\documentclass{statsoc}

\usepackage[a4paper,inner=2cm,outer=2cm, top=2cm, bottom=2cm]{geometry}
\usepackage{graphicx}
\usepackage[textwidth=8em,textsize=small]{todonotes}
\usepackage{amsmath}
\usepackage{natbib}
\usepackage{amssymb}
\usepackage{url}
\usepackage{threeparttable}
\usepackage{lineno}
\usepackage{lscape}
\usepackage{color}
\usepackage{soul}
\usepackage{comment} 
\usepackage{breqn} 
\usepackage{tikz}
\usepackage{listings}
 
\usepackage{fancyvrb}
\definecolor{myblue}{RGB}{239,245,248}
\usepackage{qtree} 
\usepackage{pdflscape} 
\usepackage{float} 
\usepackage{soul} 

\usepackage[subpreambles=true]{standalone}

\definecolor{dkgreen}{rgb}{0,0.6,0}
\definecolor{gray}{rgb}{0.5,0.5,0.5}
\definecolor{mauve}{rgb}{0.58,0,0.82} 

\newcommand{\E}{\mathrm{E}}
\newcommand{\V}{\mathrm{Var}}

\title[Correcting misclassification errors in crowdsourced ecological data: A Bayesian perspective.] {Correcting misclassification errors in crowdsourced ecological data: a Bayesian perspective.}  

\author[E Santos-Fernandez {\it et al.}]{Edgar Santos-Fernandez} 
\address{} 
\email{santosfe@qut.edu.au, edgar.santosfdez@gmail.com}
\author{Erin E. Peterson}
\author{Julie Vercelloni}
\author{Em Rushworth}
\author[]{Kerrie Mengersen \\ E-mail: k.mengersen@qut.edu.au} 
\address{School of Mathematical Sciences. Queensland University of Technology.}
\address{Australian Research Council Centre of Excellence for Mathematical and Statistical Frontiers (ACEMS)}

\begin{document}
\begin{abstract}
Many research domains use data elicited from ``citizen scientists'' when a direct measure of a process is expensive or infeasible.
However, participants may report incorrect estimates or classifications due to their lack of skill. 
We demonstrate how Bayesian hierarchical models can be used to learn about latent variables of interest, while accounting for the participants' abilities. The model is described in the context of an ecological application  that involves crowdsourced classifications of georeferenced coral-reef images from the Great Barrier Reef, Australia.  
The latent variable of interest is the proportion of coral cover, which is a common indicator of coral reef health. The participants' abilities are expressed in terms of sensitivity and specificity of a correctly classified set of points on the images. 
The model also incorporates a spatial component, which allows prediction of the latent variable in locations that have not been surveyed. 		
We show that the model outperforms traditional weighted-regression approaches used to account for uncertainty in citizen science data. Our approach produces more accurate regression coefficients and provides a better characterization of the latent process of interest. This new method is implemented in the probabilistic programming language Stan and can be applied to a wide number of problems that rely on uncertain citizen science data.
\end{abstract}


\section {Introduction}
\label{sec:fs} 

Over the last decades, citizen science (CS) and crowdsourcing projects have become
increasingly popular in several domains for tasks that require a large number of participants (i.e. volunteers or workers) to collect scientific data, generally using e-platform services \citep{bonney2014next}. 
These programs help overcome traditional scientific limitations by increasing the volume of data collected or processed, 
while also engaging the general population in science; creating awareness and helping to reach global milestones
such as the UN Sustainable Development Goals \citep{jordan2011knowledge, hsu2014development, marshall2012coralwatch}.   
Hundreds of CS projects can be found, for example, at the Federal Crowdsourcing and Citizen Science Toolkit \citep{CitizenScienceToolkit}, 
Zooniverse (\url{https://www.zooniverse.org}), iNaturalist (\url{https://www.inaturalist.org} and 
eBird \citep{sullivan2009ebird}. 
These platforms connect millions of collaborators all over the world. However, one of the main concerns when making statistical inferences using data
obtained via crowdsourcing is the inherent presence of
misclassification or measurement errors resulting from participants'
variable skill levels and abilities \citep{bachrach2012grade, venanzi2014community,
  mengersen2017modelling, claremaking2018, bird2014statistical}. 
  A second concern relates to spatial dependence in the data, which has been found to produce incorrect estimates in species abundance models when it is not accounted for \citep[e.g.][]{lichstein2002spatial, Dormann2007methods}. Spatial autocorrelation occurs naturally in many ecological datasets \citep{ver2018spatial} and this is especially true in data collected by citizen scientists \citep{fritz2019citizen}, who tend to capture observations in easily accessible areas \citep{mengersen2017modelling}. The models developed in this study address both of these important issues.

In the ecological and environmental areas, a large body of research focuses on the estimation of unbiased species abundance and distribution relative to predictors such as habitat conditions and availability, as well as anthropogenic disturbances such as the presence of roads \citep{aarts2012comparative, fithian2015bias, guelat2018effects}.
Generalized linear models (GLM) and its
variants are commonly used to assess whether one or more predictors are associated with a response variable \citep{gelfand2005modelling, bolker2009generalized}.  
However, misclassification errors in the observed variables produce biased
regression coefficients and poor model estimates, which can
substantially attenuate the influence of predictors in the model leading to potentially inaccurate
inferences \citep{fuller2009measurement,
  muff2015bayesian}.  
This issue is illustrated using a GLM with a beta distributed response
variable via simulations in the supporting web materials section. 

Approaches that pool or integrate CS elicited data with those obtained from professional monitoring programs are gathering momentum \citep{peterson2020monitoring}. 
This idea revolves around meta-analysis principles and has been extensively studied in many areas including medicine, social sciences, and the environment \citep[see][]{higgins2009re, claggett2014meta, rice2018re}, as well as ecological settings \citep{koricheva2013handbook}. There are three general approaches used to model error-prone ecological data. In the first case, measurement error is ignored, but this approach
generally produces poor estimates and is not suitable for citizen
science data that is generally messy; see the supporting materials. 
The second approach is to use weighted linear regression (i.e. weighted approach) with observation weights proportional to the user's accuracy or performance measures \citep{bird2014statistical, peterson2020monitoring}.  
This ensures that data from users with poorer accuracy receive lower weights in the regression model.  
The weighted approach may be based on fixed mechanistic weights with no variability around the values or distributions may be imposed on the weights within a Bayesian framework to represent individual accuracies. 
Generally these weights are obtained using a gold standard (e.g. expert classification) or a testing dataset where the true classes/categories are known. 
A third method  considers imperfect classification taking into consideration the user's sensitivity
and specificity, denoted here as $se$ and $sp$ respectively. In this
context, these measures refer to the ability of the citizen to
correctly detect the presence (i.e. true positive) and absence
(i.e. true negative) of the target species. See for example, \citet{petracca2018robust}.

Several variations of the second and third approaches described above rely on the Bayesian paradigm which provides a substantial number of practical and theoretical benefits \citep{bernardo2009bayesian, hobbs2015bayesian}.  
For example, a fully Bayesian formulation of the weighted approach would include a
prior distribution on the weights to capture the corresponding uncertainty of these quantities,
which would, in turn, influence the other parameter estimates and
associated uncertainty in the model \citep{choy2009elicitation}. The estimation of these users' performance measures is relevant for many crowdsourced and citizen science projects where the users are trained, and potentially compensated or rewarded for their engagement, dedication and contributions \citep[e.g.] []{amazon, wiggins2015surveying, garriga2017bayesian}.

Within the occupancy modelling framework, several authors have approached the issue of bias correction
by means of performance measures, especially the false-positive rates \citep{chambert2015modeling, claremaking2018}. 
A recent extension suggested by \citet{pacifici2017integrating} also includes a spatial component in the form of a multivariate conditional
autoregressive (MVCAR) prior, which accounts for spatial dependency in the data. 
In another example, \citet{guelat2018effects} described residual spatial autocorrelation in the context of species distribution modelling in the presence of misclassifications.
Several models that account for spatial dependency have also been developed within the Bayesian framework for citizen science data \citep{humphreys2019seasonal}.
Conditional autoregressive  (CAR) priors have been found to adequately capture spatial variability in some studies
\citep{pagel2014quantifying, purse2015landscape,arab2016spatio, arab2015spatio}, while  
Gaussian random fields \citep{humphreys2019seasonal} and stochastic partial differential equations (SPDE) \citep{peterson2020monitoring} have been successfully used in others.
To our knowledge, no one has addressed the issue of misclassification in citizen science accounting for spatial dependence.

We present a new method used to account for bias and
uncertainty in crowdsourced and citizen science data, which we refer to as a spatially
dependent misclassification error (SDME) approach. 
More specifically, we propose a Bayesian hierarchical spatial model  
used to correct misclassification errors and estimate the latent proportion of the target region occupied by a given species or ecological community.
In the context of our case study, we are interested in estimating the percent or proportion of the seafloor covered in hard corals within an area; hereafter referred to as hard coral cover. We demonstrate the approach using simulated data and field estimates of coral cover collected within the Australian Great Barrier Reef between 2008-2017. Finally, we discuss the advantages
and disadvantages of the approach, as well as the main implications of
the research. 
In keeping with the case study, we concentrate on the classification of images, 
but the methods are also applicable to
the analysis of other sources such as text, video, audio, etc. found in citizen science.

\subsection{Motivating dataset: classification of coral reef images}
\label{sec:motiv} 

\subsubsection{Origins of coral reef images}

The Great Barrier Reef (GBR) is one of the richest and most complex ecosystems in the world. However, coral reefs are negatively affected by  pressures such as climate change, 
which have caused a substantial decline in hard coral abundance \citep{de201227, ainsworth2016climate, authority2014great}. In addition, monitoring the GBR is especially challenging because it extends over 346,000 km$^2$ and traditional marine surveys are expensive \citep{nichols2006monitoring, roelfsema2010calibration, nygaard2016price, Vercelloni15059}. 
To address these issues, \citet{peterson2020monitoring} demonstrated how image-based coral cover data elicited from citizens can be combined with other professional data sources to improve the spatio-temporal data coverage in the GBR and increase the information gained to inform management. This approach was operationalised in the Virtual Reef Diver project using the weighted approach (\url{https://www.virtualreef.org.au/}), without accounting for misclassification bias. Therefore, we developed an experiment within the Virtual Reef Diver project to further assess the potential of citizen science and crowdsourced data for monitoring coral cover. Our main purpose is to assess the participants' abilities to identify hard corals within geotagged images while determining the impact of different reef disturbances on hard coral cover changes. In addition, we want to evaluate the quality of the estimates we obtain from the experiment.

We obtained a set of $N = 1585$ images from 
unique locations across the GBR taken between years 2008 and 2017 by
the XL Catlin Seaview Survey \citep{gonzalez2014catlin} and
the University of Queensland's Remote Sensing Research Centre \citep{roelfsema2018coral}, which had been previously classified by marine scientists.
Each image contained 40 spatially balanced, random classification points. 
It is common to treat classifications from coral reef scientists as the gold standard, with no uncertainty associated with the measurements. However, ecologists are often
interested in data from large regions where it is infeasible to obtain
estimates of the proportion of hard corals ($y_j$) for large volumes of images due to time and financial constraints. In these cases, citizen science and crowdsourcing
are excellent alternatives \citep{dickinson2010citizen}.

\subsubsection{Spatial covariates for coral cover}

Four spatial covariates representing reef disturbances, management
zones, and continental reef-shelf position were considered in the model (Table \ref{table:predic}). 
Reef disturbance covariates were sourced from \citet{matthews2019high}.
The maximum value of Degree Heating Weeks (DHW) is generally used to represent a reef's thermal stress and has been found to be associated with episodes of coral bleaching \citep{hughes2018spatial}.  
Exposure to cyclones is defined as the number of hours with potentially damaging waves (height $>$ 4m) during tropical cyclones or storms \citep{puotinen2016robust}. 
Note that, nearest neighbour interpolation was used to fill in locations where covariate values were missing.

\begin{table}
\caption{\label{table:predic} Covariates included in the model.}
\scalebox{0.850}{
\begin{tabular}{llll} 
\hline
Covariate	& Variable& Description & Source\\ \hline
DHW	&continuous & Degree heating weeks (max DHW/year).& \citet{hughes2018spatial, matthews2019high}\\
  	&  & It is a proxy for bleaching severity. & \\
no\_take&	binary \{no-take = 1, & No-take marine reserves where  & \citet{GBRMPA2014} \\
&	  take = 0\} & fishing is not allowed. & \\
shelf	&	categorical \{middle = 0, & Position of the reef. For these  & \citet{GBRMPA2014}\\
&	 outer = 1\} &  images we had no inner reefs.  & \\
CYC	  &	continuous & Cyclone effect 
measured as cumulative hours \\
	  &	 & of exposure to waves greater than 4m (4MWh/year). & \citet{puotinen2016robust, matthews2019high}\\

\hline
\end{tabular}
}
\end{table}

\begin{figure}[htbp]
	\centering
		\includegraphics[width=4.50in]{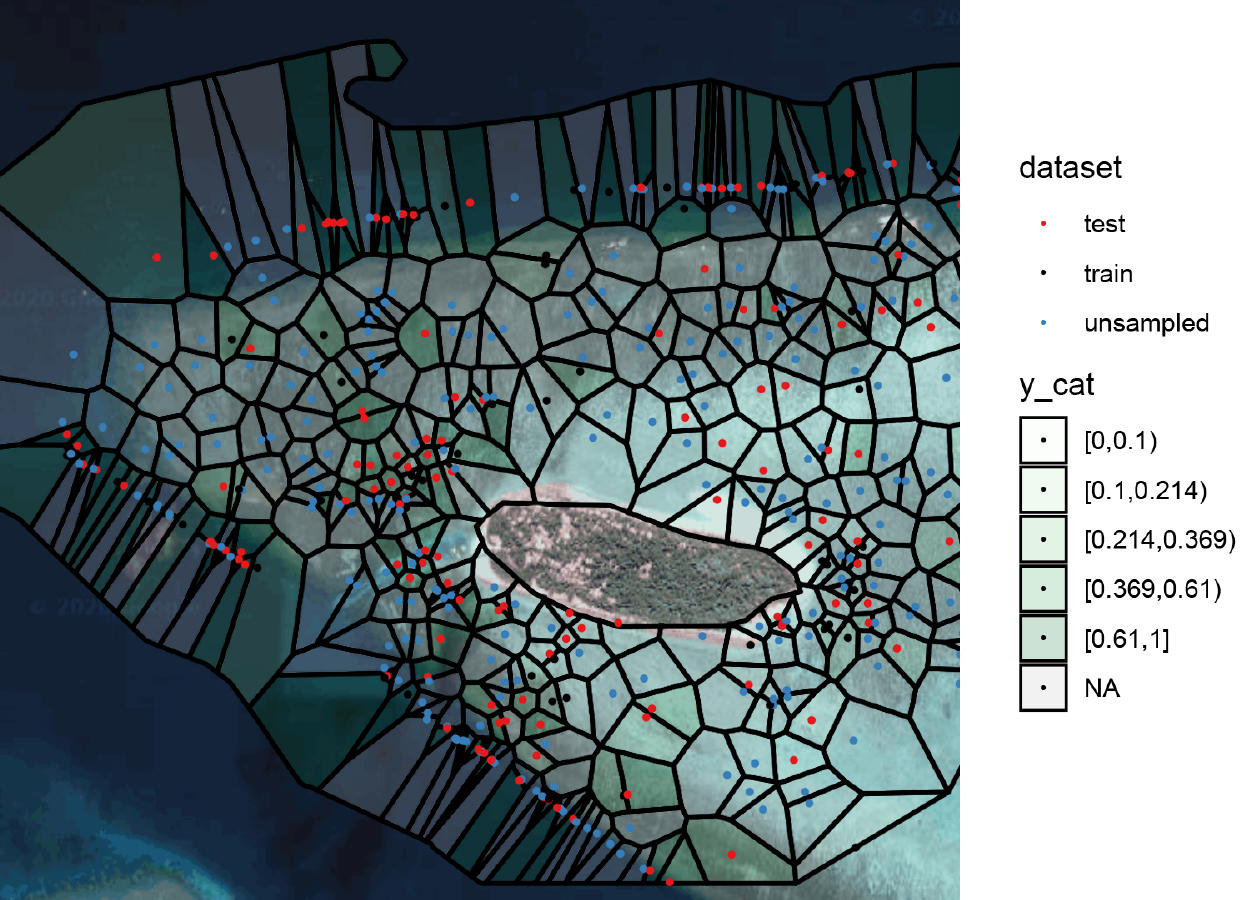} 
    \caption{Voronoi diagram of hard coral cover around Heron
    Island, Australia. The categorical colors are the
                  classes obtained using the quintiles of the hard
                  coral proportion obtained by expert elicitation.
                  The dots represent the location of the
                  measurement.
	Black and red dots are locations where images where taken and in images given by the black dots the latent true proportion is available. Blue dots in Voronoi areas are positions on the reef in which we want to predict the coral cover.   }
	\label{fig:1Fig1a}
\end{figure}

We assume that images in close proximity within the same reef habitat are more likely to have a similar proportion of hard coral
than those situated in a different habitat, motivating the use of a spatial component in our models. Spatial autoregressive models \citep{ver2018spatial} or SPDEs (Lindgren and Lindgren 2011) are often used to model large datasets because they are more computationally efficient than geostatistical models applied to spatially continuous point-referenced data (Cressie and Wikle 2011). 
Therefore, we generated a set of Voronoi polygons using the image locations as centroids and boundaries based on Euclidean distance \citep{okabe2009spatial, gold2016spatial}, which we then used to define the spatial domain in the model. Two polygons are neighbors if they share a common boundary.
The example in Fig \ref{fig:1Fig1a} shows a Voronoi diagram with the spatial proportion of hard coral on the edges of a reef from Heron Island, Australia. This diagram is superimposed on a background satellite image obtained from Google. The black and red and points represent the locations
where images were taken, while the blue ones represent locations of interest for reef management
purposes where an estimate of the proportion of hard corals are required. The images capture an extent over a certain area of the reef defined by transects.  

\section{Materials and methods}
\label{sec:mm}

\subsection{Survey design and description of the classification task}

For each image, we obtained the proportion of five benthic categories (i.e. hard and soft coral, algae, sand, and other). 
The images were clustered into groups based on proportions of each of the five categories using the K-Means clustering algorithm 
included in the \emph{stats} package in R.
This approach produced three clusters of images based on their benthic composition: (1) images mostly composed of hard corals, (2) largely dominated by algae, and (3) predominantly composed of soft corals.

Then, a random sample of 514 images was selected from the clusters ensuring that we had coverage across the camera types used to take the images, which had different resolutions (Canon, Lumix, Olympus, Sony and Nikon). This design ensured that the images represented the benthic composition and camera types found in the full set of images, as well as a wide spectrum of classification difficulties. 
We randomly selected 30\% of the 514 images ($n = 171$) and assumed the corresponding proportion of hard corals $y$ to be known (i.e. training dataset) as the result of elicitation by the coral reef scientists.
For the remaining 343 images (i.e. testing dataset), we assumed the proportion of hard corals was unknown and must be estimated from the model based on the participants' responses. This combined dataset of 171 known values of $y$ and 343 estimated values of $y$ was then used to estimate the model parameters.
Finally, we treated the 1071 remaining image locations as an unsampled/prediction dataset, which allowed us to predict coral cover using the fitted model and also validate the model results.

There exists a trade-off between survey costs and the quality of the estimates (e.g. participants' performance measures) and knowing 30\% of the labels is generally considered suitable for these kind of problems. 
This and other training partitions have been found suitable in the literature. See e.g. \citet{pacifici2017integrating} who consider 25\% and 50\% values, while \citet{chambert2018two} deal with smaller proportions such as 5, 10, 20 and 30\%.

The 514 images were displayed for classification on Amazon Mechanical Turk (\url{https://www.mturk.com/}) and workers were asked to classify points into five benthic categories (i.e. hard coral, soft coral, algae, sand, and other). 
We created a help file showing underwater image classification to train the participants, which described the characteristics of the benthic categories (\url{https://github.com/EdgarSantos-Fernandez/reef_misclassification/blob/master/HelpGuide_MTurk2020200203.pdf}).   
Participants also had to pass a qualification test by achieving a score of at least 60\% to perform classifications, which is a common mechanism of quality control \citep{rashtchian2010collecting}. Classification data were collected from 2020-Jan-14 to 2020-Feb-12.

Participants were paid 0.10 USD per completed image, which equates to more than the U.S. federal mininum wage (\$7.25 per hour).
The number of images classified by each participant varied because they were free to cease the task at any time. 
Participants were assigned a sequence of images from the list of 514 obtained by random sampling without replacement.

\begin{figure}[htbp]
	\centering
     \includegraphics[width=4.25in]{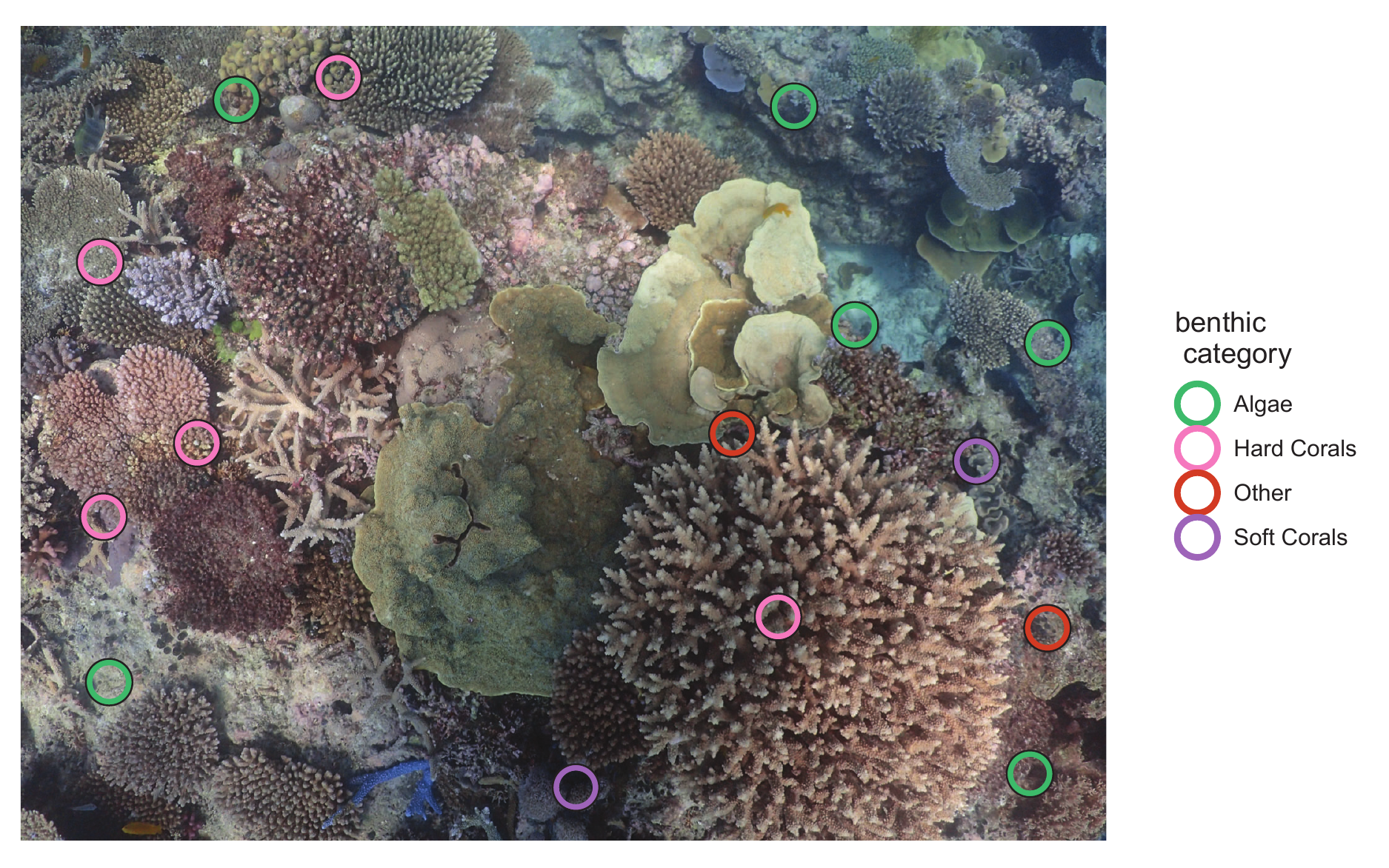} 
     \caption{Underwater image from 
		Heron Island Reef
       Management Area, Great Barrier Reef, Australia. 
			It shows 15 spatially balanced random points elicited by a marine scientist. The color of the circle represents the true
       benthic category, which is defined as the class with the highest
       proportion within the area delimited by the circle. 
     } 
    \label{fig:2Fig2}
\end{figure}

For each image, they were asked to classify 15 (out of the 40) random points and the task could only be submitted after classifying all the points on an image. 
The structure of the data will be discussed below in Fig \ref{fig:plot_diagram}. 
We considered asking participants to annotate the full 40 points per image, but we were concerned that it would be too exhausting for the participants and would reduce participation.  
In addition, \citet{beijbom2015towards} showed that accurate image-based estimates of coral cover could be obtained with approximately 10 points per image, and that manual classification of additional points by marine scientists did not substantially improve coral cover estimates for an individual image.
Fig.~\ref{fig:2Fig2} shows one of the underwater images taken from Heron Island Reef.
The true classes are shown for the 15 spatially balanced random points classified by a participant. 

Once data were submitted by workers, some exclusion rules were put in place to discard non-informative data, 
careless and non-genuine users, and software bots.
This was done by using the values of coral cover from the training dataset. Subjects with accuracy values lower than 40\% in the training images were excluded. 
We also considered other indicators including the number of classifications per hour and inconsistencies on some of the fields of the database tuples.
Failing to remove these noisy data points generally results in biased estimates.
Thus, the final dataset used for the case study comprised classifications from 157 subjects and included 212,910 observations (i.e. classification points) and 14,194 image classifications.

\begin{figure}[hptb]
	\centering
		\includegraphics[width=6.5in]{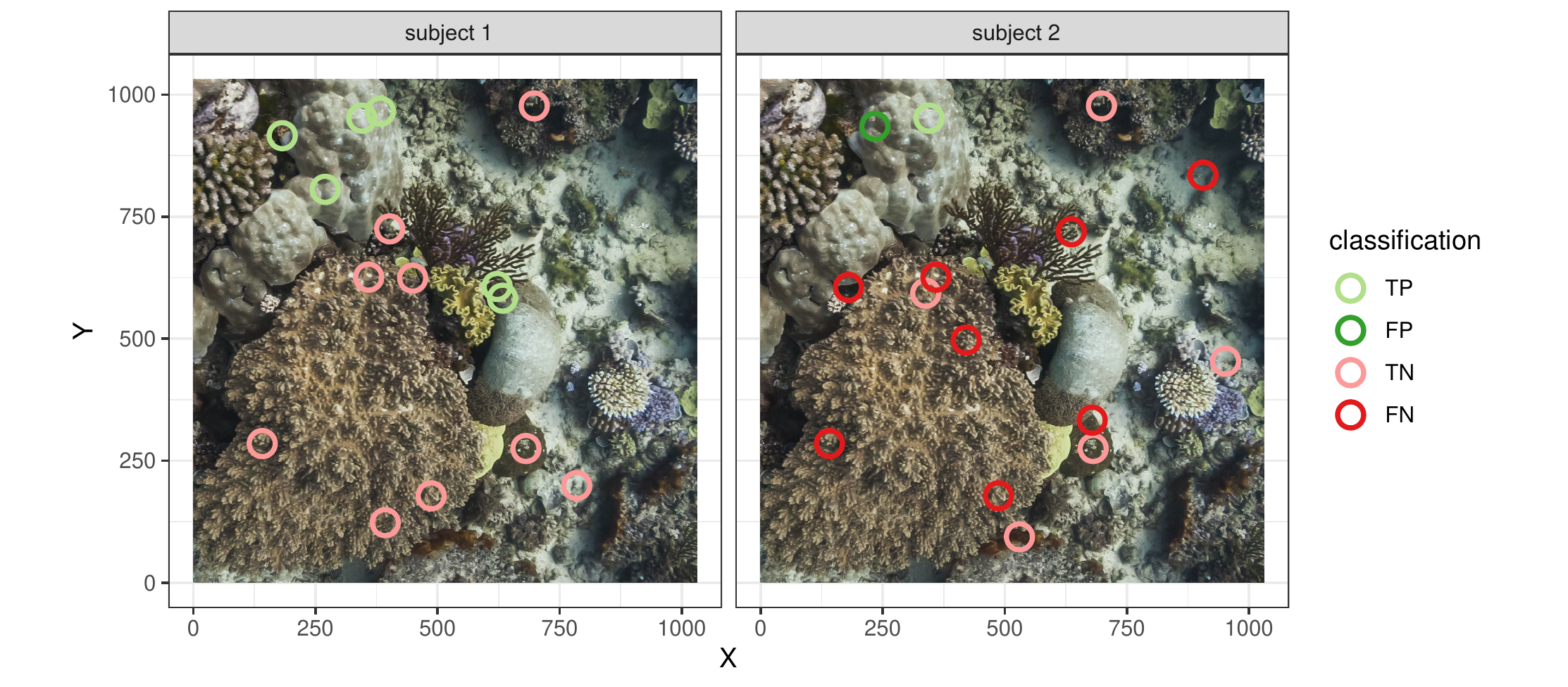}
                \caption{Example of an image classified by two
                  subjects with different abilities to classify corals. Each image
                  contains $q = 15$ random classification points that
                  are colour coded to represent whether hard coral has
                  been classified correctly (true positive). Dark colours represent
                  misclassification (dark green = false positive (FP),
                  dark red = false negative (FN)). Light colours represent correct classifications (light green = true positive (TP), light red = true negative (TN)).}
\label{fig:3Fig3}
\end{figure}

Fig.~\ref{fig:3Fig3} shows an example of the elicitations from two subjects with different abilities to classify hard coral. We found that images with a large proportion of algae tended to produce larger false-positive rates compared to (easier) images with a large portion of sand. This is probably because algae look like hard corals to less proficient subjects.


\subsection{Statistical description of the classification task}
In this section, we provide a statistical description of the classification task, the misclassification errors and accuracy, and the spatial Bayesian hierarchical model. A complete list of symbols and definitions has been provided in the Appendix section that readers can refer to. 

Consider coral reef images taken at geographic locations defined by latitude and longitude (lat and lon). Let these locations represent the centroids of Voroni polygons used to define areal units. 

The presence or absence of the target class (hard corals) in a subset of points is obtained within each image. 
We denote $y$ as the true proportion of these points containing the target class. 

In binary image-classification tasks, participants are asked whether each of $q$ sample points contain the class of interest. 
In the case study described above, $q=15$ and the target class is hard corals.
This approach is also known in ecology as random point count methodology \citep{kohler2006coral}. 
Let $z_{ijk} = \{0,1\}$ with ``1'' denoting the target class, in the point $k$ from the $j^{th}$ image classified by the $i^{\textrm{th}}$ subject. 
Thus, there might be a disagreement between the elicited and the true latent class.
For a given image $j$, the \emph{apparent} proportion of the target species
($\hat{y}_{ij}$) is obtained as the sum of points that are labeled as ``1''
divided by $q$.

\begin{equation} 
\hat{y}_{ij} = \sum_{k=1}^{q} z_{ijk}/q
\end{equation} 

\subsection{Characterization of misclassification errors and participant's accuracy}
 
The performance of the $i^{th}$ subject is measured by their sensitivity ($se_i$),
 specificity ($sp_i$) and accuracy($acc_i$), which are obtained as follows:  

\begin{equation}
se_i = \frac{\sum_{j=1}^{m}\sum_{k=1}^{q}TP_{ijk}}{\sum_{j=1}^{m}\sum_{k=1}^{q}TP_{ijk} + \sum_{j=1}^{m}\sum_{k=1}^{q}FN_{ijk}}, 
\label {eq:eqse}
\end{equation} 

 \begin{equation}
 sp_i = \frac{\sum_{j=1}^{m}\sum_{k=1}^{q}TN_{ijk}}{\sum_{j=1}^{m}\sum_{k=1}^{q}TN_{ijk} + \sum_{j=1}^{m}\sum_{k=1}^{q}FP_{ijk}} ,
\label {eq:eqsp}
\end{equation} 

and 

\begin{equation}
 acc_i = \frac{\sum_{j=1}^{m}\sum_{k=1}^{q}TP_{ijk} + \sum_{j=1}^{m}\sum_{k=1}^{q}TN_{ijk}}{\sum_{j=1}^{m}\sum_{k=1}^{q}TP_{ijk} + \sum_{j=1}^{m}\sum_{k=1}^{q}FN_{ijk}+ \sum_{j=1}^{m}\sum_{k=1}^{q}TN_{ijk} + \sum_{j=1}^{m}\sum_{k=1}^{q}FP_{ijk} },
\label{eq:eqacc}
\end{equation}

\noindent 
where for the $k^{th}$ point on the $j^{th}$ image classified by the subject $i$, 
the four indicator variables $TP_{ijk}, TN_{ijk}, FP_{ijk}, FN_{ijk}$ are equal to $0$ or $1$,
with $TP_{ijk} = 1$ if the point is correctly classified as positive given that the target species is present (true positive) and
$TN_{ijk} = 1$ if the point is correctly classified as negative when the target species is absent (true negative). 
The false positive $FP_{ijk} = 1$ when the point is incorrectly classified as positive when the target species is absent and
the false negative $FN_{ijk} = 1$ occurs when it is misclassified as not present when the target species is present in the location. For a discussion on the 2$\times$2 confusion matrix in the ecological context see \citet{manel2001evaluating, vayssieres2000classification}.

Accounting for a proportion of locations where the $y_j$ is known, informs the
model as a training set about the users' abilities in terms of $se_i$ and $sp_i$. 
This will narrow the uncertainty around the latent values in locations where the truth is unknown and
contribute to model identifiability.

We will see later in Eq.~\ref{eq:yhat} that the participant's performance plays a vital
role in the model.
Accuracy estimates from previous citizen science
studies have ranged between 70 and
95\% \citep{kosmala2016assessing}, with a subject's classification performance affected by commitment, effort, ability and selected demographic factors.  

Within the Bayesian framework, $se_i$ and $sp_i$ for each participant are assumed to follow a probability 
distribution with parameters that reflect published estimates. 
Other potential cases consider distributions for each participant $i$ and each image $j$ ($se_{ij}$ and $sp_{ij}$) or
consider $se_i$ and $sp_i$ to be a mixture of distributions for easy and hard images for example.
Here we adopt informed beta distributions with relatively large shape
$\alpha_i$ and small scale $\beta_i$, which produces a density with mass closer to 1 than to 0 with a relatively small variance.

\subsection{The SDME and the weighted Bayesian hierarchical model. }
\label{sec:modeling}

\begin{figure}[htbp]
	\centering
		\includegraphics[width=3.5in]{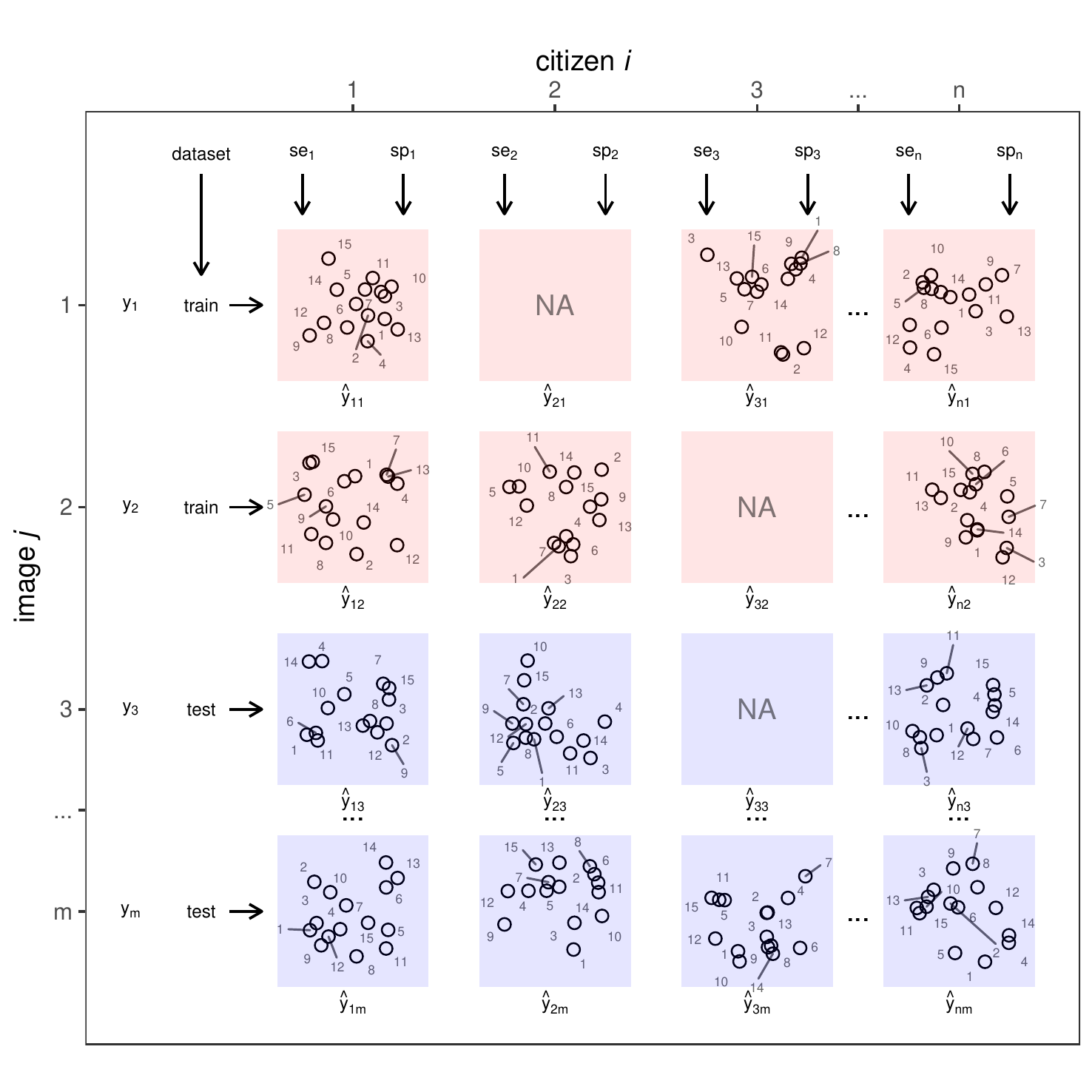}
	\caption{Structure of the data. Tiles represent the image $j$ (rows) being classified by the subject $i$ (columns) with $se_i$ and $sp_i$.
	On each classification 15 points are elicited producing estimates of $\hat{y}_{ij}$.
	Images in red are those used for training while the blue ones give the testing subset. 
	A tiles with NA means that the image $j$ was not classified by the subject $i$.}
	\label{fig:plot_diagram} 
\end{figure}

Participants with limited training may find it difficult to
correctly identify some of the benthic categories found at the sample points.
The statistic $\hat{y}_{ij}$ gives the apparent proportion of hard corals in the images $j = 1, 2, \cdots, m$ classified by the subjects $i = 1, 2, \cdots, n$ and it is obtained deterministically based on the subject $se_i$ and $sp_i$ distributions and the true latent proportion $y_j$ according to Eq \ref{eq:yhat} \citep[e.g.][]{vose2008risk}. 
A graphical representation of the image classification process is given in Fig \ref{fig:plot_diagram}.

\begin{equation} 
\hat{y}_{ij} = y_{j} \times se_{i} + (1 - y_{j})\times(1 - sp_{i})
\label{eq:yhat}
\end{equation} 

\begin{figure}[htbp]
	\centering
		\includegraphics[width=4.0in]{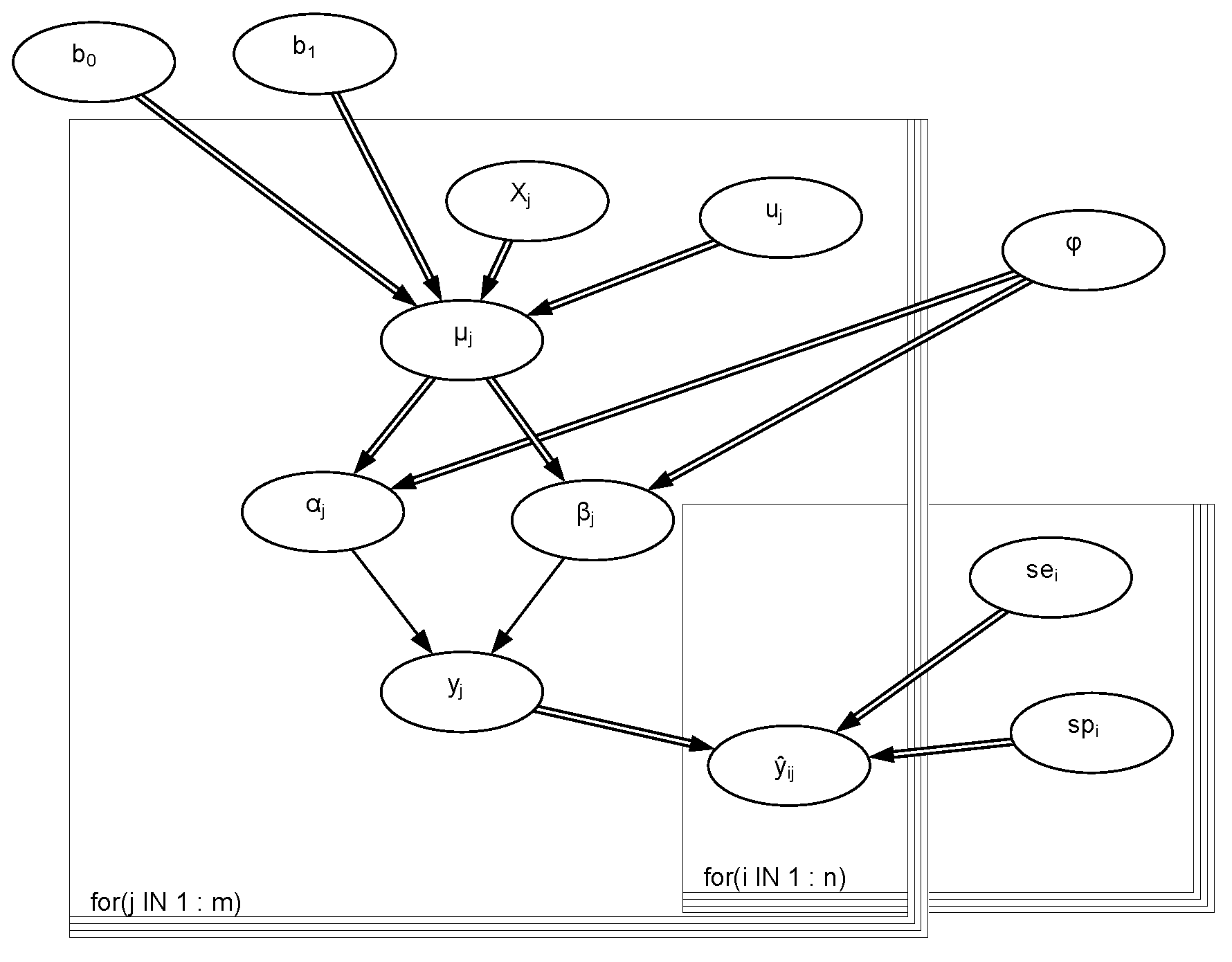}
	\caption{Directed acyclic graph of the SDME model. Ellipses represent the stochastic/constant nodes. 
	Solid and hollow arrows (edges) give stochastic and logical dependence, respectively.
	Plates represent the repeated elements. 
	The left plate is indexed in $j$ representing the portion of the model relating to images, while the portion relating to the workers is shown in the right plate and is indexed in $i$.
	For example, $se_{i}$ is indexed in $i$ (i IN 1 : n) indicating
	the sensitivity of users from 1 to $n$.  
	} 
	\label{fig:DAG}
\end{figure}

The SDME model is depicted in the directed acyclic graph in Fig.~\ref{fig:DAG}, which contains two plates: one for the image part ($j$) and the other representing the subjects ($i$). 

When the target class of interest is absent at all of the points in the
 $j^{\textrm{th}}$ image ($y_j = 0$) ,
 the apparent proportion $y_{ij}$ is the false positive rate of the subject $i$, i.e., $y_{ij} = 1-sp_i$.
If participant $i$ has a small $sp_{i}$, the participant will report larger values $\hat{y}_{ij}$ for small values of $y_j$.
When the target class is present in all the points ($y_j = 1$), the apparent proportion will be equal to the subject's sensitivity $\hat{y}_{ij} = se_{i}$.

The true proportion of the target class $y_j$ within the image $j$ is
modelled in our case using a beta regression approach \citep{ferrari2004beta}, which is a common practice in reef modelling \citep[e.g.][]{mcclanahan2019temperature, mellin2019cross, peterson2020monitoring}:

$$y_j | \alpha_j, \beta_j  \sim \textrm{Beta}(\alpha_j,\beta_j )$$

\noindent where $\alpha_j$ and $\beta_j$ are the shape and the scale parameters respectively. 
We parametrized the model based on the mean $\mu_j$ and a common precision parameter $\phi$, where 
$\alpha_j = \mu_j \phi$ and $\beta_j = -\mu_j \phi + \phi$. 
The mean is conditional on the latent model, $\mu_{j} = \E[y_j|\alpha_j, \beta_j]$, and the variance
$\V(y_j) = \frac{\mu_{j}(1-\mu_{j})}{(1+\phi)}$, 
where $\phi$ is inversely proportional to the variance of $y_j$.

The spatial beta regression for each image/unique location $j = 1, 2,\cdots, J$ can be expressed as follows:

\begin{equation} 
\textrm{logit}(\mu_j) = X_jb + u_j + \varepsilon_j,
\label{eq:reg}
\end{equation}

\noindent where the matrix $X_j$ represents a group of covariates, $b$
is the vector of regression coefficients,
 $\varepsilon_j$ is unstructured noise 
and $u_j$ is a spatial
component obtained from a spatial autoregressive model such as the
conditional autoregressive (CAR) model prior \citep{besag1991, ver2018spatial}. 
In a CAR model, the value at $u_{l=1,2,\cdots,m}$ conditional on the first-order neighbours is the average of the $n_l$ first-order neighbours plus Gaussian noise.

\begin{equation}
u_{l}|u_{t},\tau_u \sim \mathcal N \left ( \frac{1}{n_l}\sum_{l \sim t}^{.}u_t, \frac{1}{\tau_u n_l} \right ),
\label {eq:eq3001}
\end{equation}
 
\noindent where $l \sim t$ means $l$ and $t$ are neighbours and $l \neq t$.
This prior has been implemented in Stan \citep{MORRIS2019100301}.

\begin{figure}[htbp]
	\centering
		\includegraphics[width=4.0in]{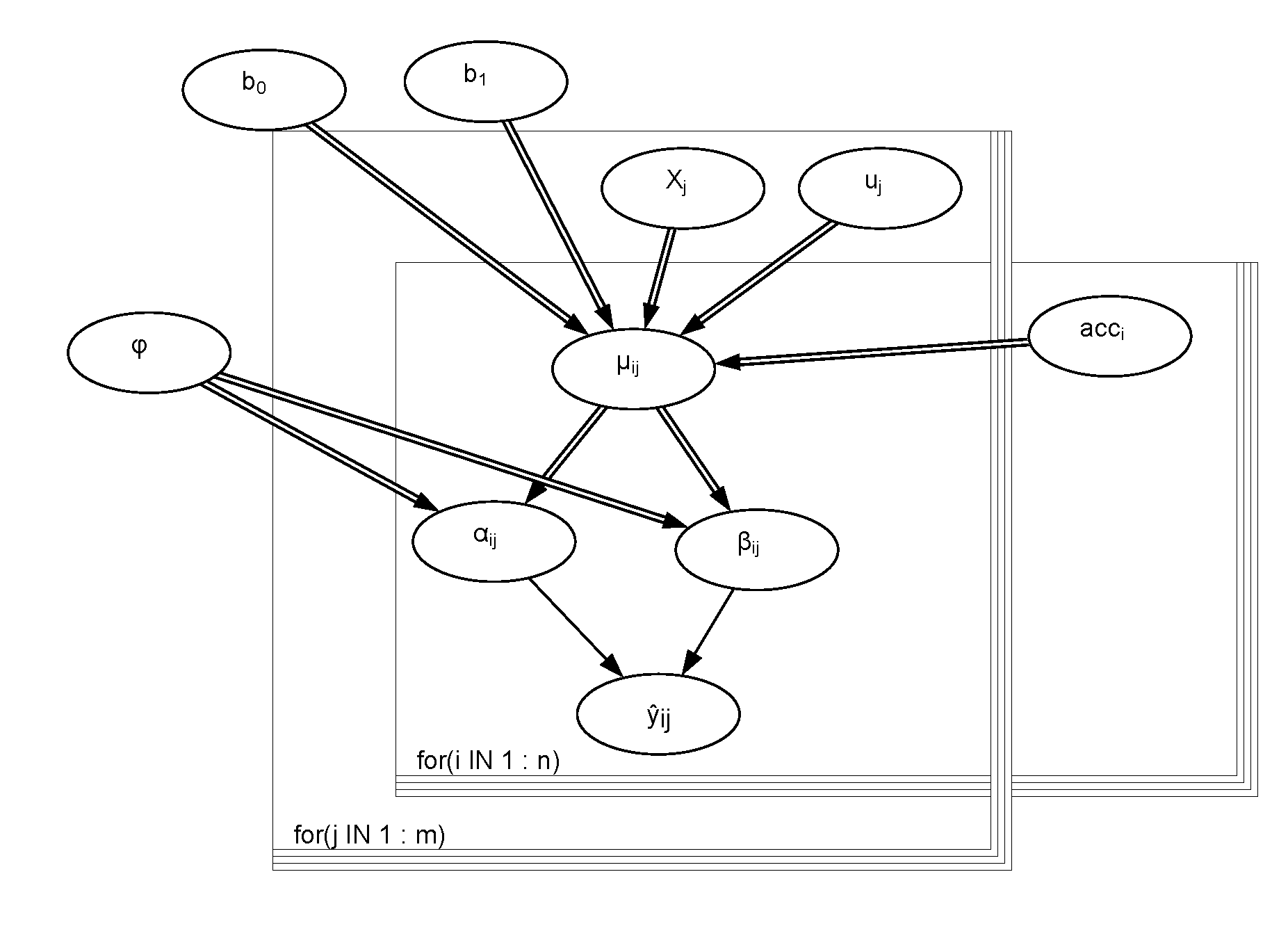}
	\caption{Directed acyclic graph of the weighted model. 
	} 
	\label{fig:DAG_w}
\end{figure}

The Bayesian weighted model is shown in Fig.~\ref{fig:DAG_w}. 
In this model, the node $acc_i$ is the
subject $i$ classification accuracy defined above in Eq. \ref{eq:eqacc}, which is used 
to weight the values of $\mu_{ij}$ in the regression model.
We model $acc_i$ using a beta prior distributions obtained from the training dataset.

\section{Simulation study}
In order to assess how well the model parameters are estimated, we simulated
300 datasets and  compared the fit using the weighted and the SDME models.
In the SDME model, the aim was to estimate the parameters from Eq. \ref{eq:yhat} and \ref{eq:reg}:
$\theta_{SDME} = \left \{ y_i, b_0, b_1, se_i, sp_i, \phi \right \}$
using the apparent proportion $\hat{y}_{ij} = (\hat{y}_{11}, \hat{y}_{12},\cdots,\hat{y}_{nm} )$, some available $y_j$ values and a covariate $X = \left[x_1, x_2, \cdots x_j\right]$. 
In the weighted model, we estimated $\theta_{w} = \left \{\hat{y}_{ij}, b_0, b_1, acc_i, \phi \right \}$.

We also wanted to obtain the posterior estimates of $y_i$ for unobserved locations $j$ where no images were taken, where $\hat{y}_{ij}$ is missing for all participants $i = 1, 2, \cdots, n$.
The simulation steps are detailed below, followed by a discussion of parameter choices.

\begin{enumerate}
	\item 
	Consider 225 points on a unit square with \emph{lat} and  \emph{lon} on a plane representing a section of a reef.
	We select 80\% of these points and take one image at each point.
	We define Voronoi cells based on Euclidean distance using the centroids defined by the \emph{lat} and \emph{lon}. 
	 
	\item 	Assume a continuous covariate 	$x_j \sim \textrm{N}\left(\mu_x = 0, \sigma_x \right)$ 
	sampled at the locations where the images were taken. 
	It represents a reef disturbance and is highly associated with $y_j$. In the context of the case study this could represent a change in water temperature (e.g. DHW).

	\item Fix an intercept $b_0 = 1$ and a slope $b_1 = -2$. These values are arbitrarily chosen for the purpose of the simulation study and other values do not seem to affect the results.
	\item Include 20 subjects from 4 groups with different levels of accuracy (1-4): experienced, good, average and beginner. 
	We assume performance measures based on the ranges reported in \citet{kosmala2016assessing} and hence randomly sample beta distributions for the $se_i$ with mean values according to the expertise membership: 
	$se_1 = 0.99$, $se_2 = 0.95$, $se_3 = 0.90$ and $se_4 = 0.80$ and using a precision $\phi_{se} = 50$ based on the beta distribution parametrization discussed in Section \ref{sec:modeling}.
	Similarly we set the mean $sp$ values: $sp_1 = 0.99$, $sp_2 = 0.90$, $sp_3 = 0.80$ and $sp_4 = 0.70$ and use a precision $\phi_{sp} = 50$. This ensures that for every subject the probability of correctly classifying each point is stochastic.

	Apart from the experienced subjects (group 1), the others have a larger false positive rate than false negative rate ($sp_i < se_i$, for $i = \{2,3,\cdots\}$). 
	Images are classified by at least 5 participants and the probability of an image being classified by 5,6,$\cdots$ or 20 participants is the same. 
	Every user classified approximately 3/4 of the total number of images. 
	In the SDME model, informative beta priors are used for $se_i$ and $sp_i$, with shape parameters calculated using the mean and variance obtained from the training dataset. 
 \item Set the beta precision parameter $\phi = 30$. Compute the
        beta distribution mean (Eq \ref{eq:reg}) based on the spatial
        regression equation. Compute also the shape and scale parameters
        ($\alpha_j$, $\beta_j$).
  \item Simulate latent responses $y_j$ for $j = 1, 2, \cdots, m$, and based on these compute the apparent responses $\hat{y}_{ij}$ for $i = 1, 2, \cdots, n$ and $j = 1, 2, \cdots, m$.

	\item There is missingness in both $y_j$ and $\hat{y}_{ij}$ (Fig \ref{fig:datasets}).
	This yields the three datasets structures that are commonly encountered in CS ecological research. 
	The first split is whether samples or images being collected at the given locations and the images have been classified (is $\hat{y}_{ij}$ available?), while the second depends on the underlying truth ($y_j$) being available.
	Define a vector $d_j = \{0,1\}$ with  $j = 1,2,\cdots,m$ where $d_j = 1$ if $y_j$ is observed and $d_j = 0$ otherwise.
	Let also $\hat{d}_j = \{0,1\}$, associated to $\hat{y}_{ij}$, where $\hat{d}_j = 1$ indicates that at least one subject has classified the image $j$.   
	
	In dataset 1 (i.e. training) both $\hat{y}_{ij}$ and $y_j$ are available and $d_j = \hat{d}_j = 1$. 
	We randomly selected 67 images (30\%) where the true proportion ($y_j$) is known with certainty.
	The second dataset (i.e. testing) is represented by those locations where images were collected and will be classified (known $\hat{y}_{ij}$, $\hat{d}_{j} = 1$), but the ground truth is not available (unknown $y_{j}$, $d_j = 0$). 
	
Finally, we consider 45 locations (20\%) that have not been sampled yet and thus there are no images, so both $\hat{y}_{ij}$ and $y_j$ are missing (i.e. unsampled dataset, $d_j = \hat{d}_j = 0$).  
We predict the proportion of the target species in these locations where no data have been collected using neighboring information and covariates.

\item For the study, we generated a Voronoi grid of size 15$\times$15 spatial locations points on a unit square ([0,1] $\times$ [0,1] ) corresponding to the positions where the images were taken.

\end{enumerate}

\begin{figure}[htbp]
	\centering
	\Tree[.dataset 
          [.\textsc{sampled locations} 
		[ .\textit{\textcolor{red}{(1) training }}
		[.\textit{(known $\hat{y}_{ij}$ $\&$ $y_j$; $d_j = \hat{d}_j = 1$}) ]]
			[.\textit{(2) testing} 
			[.\textit{(known $\hat{y}_{ij}$ and unknown $y_j$; $d_j = 0, \hat{d}_j = 1$)} ]
								]]
						[.\textsc{\textcolor{blue}{(3) unsampled locations} }
								[.\textit{(unknown $\hat{y}_{ij}$ $\&$ $y_j$; $d_j = \hat{d}_j = 0$)} ]
							]]

	\caption{Illustration of the three datasets used by the model. 
	In locations where samples (images) have been classified $\hat{y}_{ij}$ is available. 
	Dataset 1 (training) corresponds to those where the ground truth ($y_j$) is known, while the second (testing) is for the unknown $y_j$ case. 
	Unsampled dataset refers to geographical locations for which no samples or images have been collected and therefore both $\hat{y}_{ij}$ and $y_j$ are unknown. 
			} 
	\label{fig:datasets}
\end{figure}

\begin{figure}[h]
	\centering
		\includegraphics[width=6.75in]{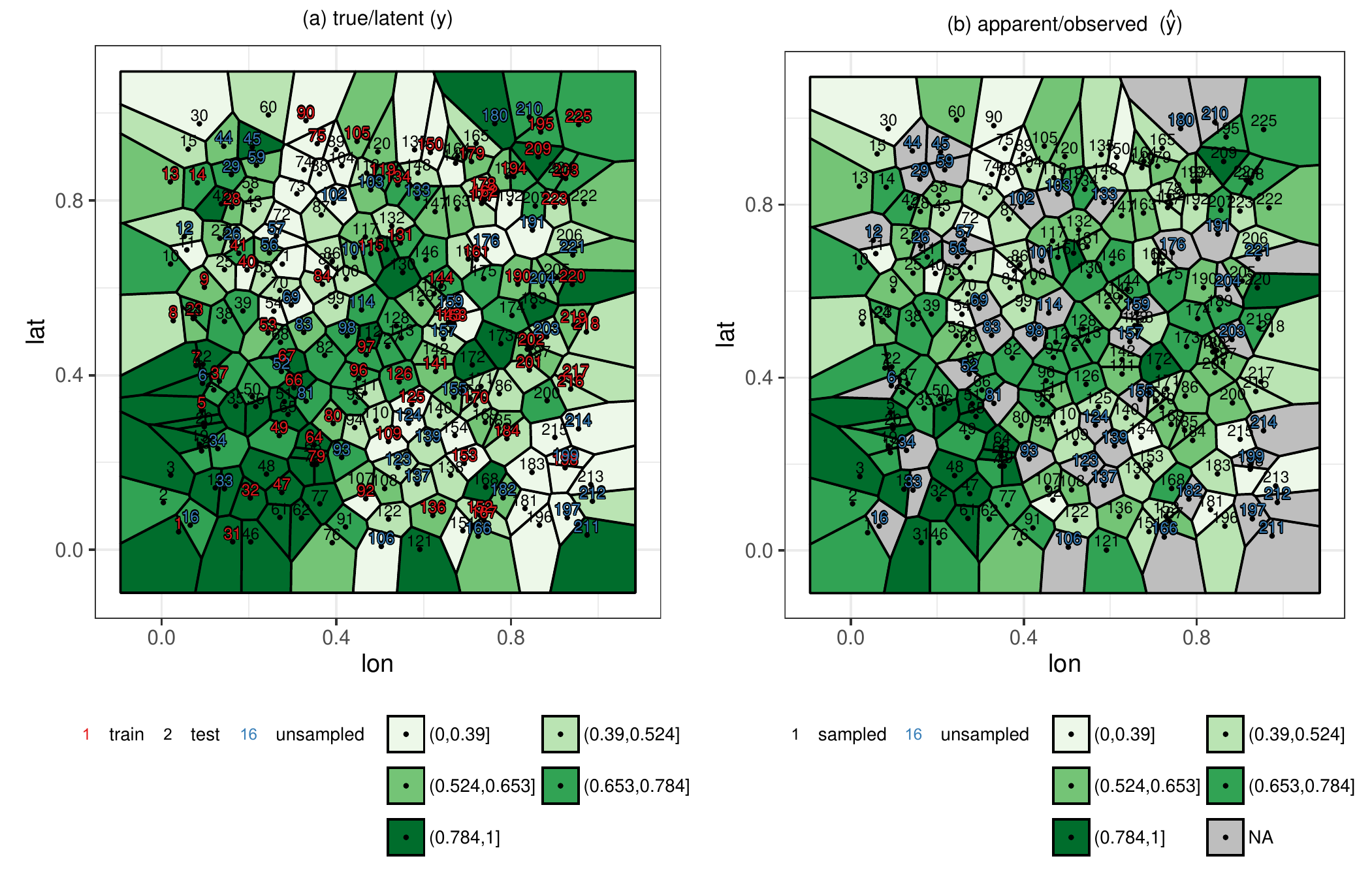}
		\vspace{-0.5cm}
                \caption{ (a) Voronoi diagram of the true latent
                  proportion ($y_j$) at 225 simulated
                  locations. 
Each Voronoi cell or polygon has a centroid represented by a point and each point is labelled from 1 to 225 in which the color represents the type of dataset (train, testing, unsampled).
The true value $y_j$ is known at
                  67 locations (red labels). The 45 polygons with blue labels are the unsampled locations. Black labels represent polygons from the testing dataset.
The green color categories are obtained using the quintiles.
(b) Voronoi diagram of the apparent proportion of coral cover ($\hat{y}_{ij}$)
                  elicited by the subjects. Missing values in
                  $\hat{y}_{ij}$ are shown in (gray polygon with text in
                  blue). There are 45 mismatches (out of 80 locations)
                  in the proportion categories in (a) and (b): 1, 2,
                  8, etc. since users mostly overestimated the
                  proportion of species.}
	\label{fig:1Fig1}
\end{figure}

The simulated proportion of hard corals at the 225 locations is shown in
Fig.~\ref{fig:1Fig1}(a) and spatial association between adjacent
areas is evident.
In the 45 unsampled locations we set $\hat{y}_{ij}$ to missing as described
above (\ref{fig:1Fig1}b; gray polygons and blue labels).

\begin{figure}[h]
	\centering
	
		\includegraphics[width=5.2in]{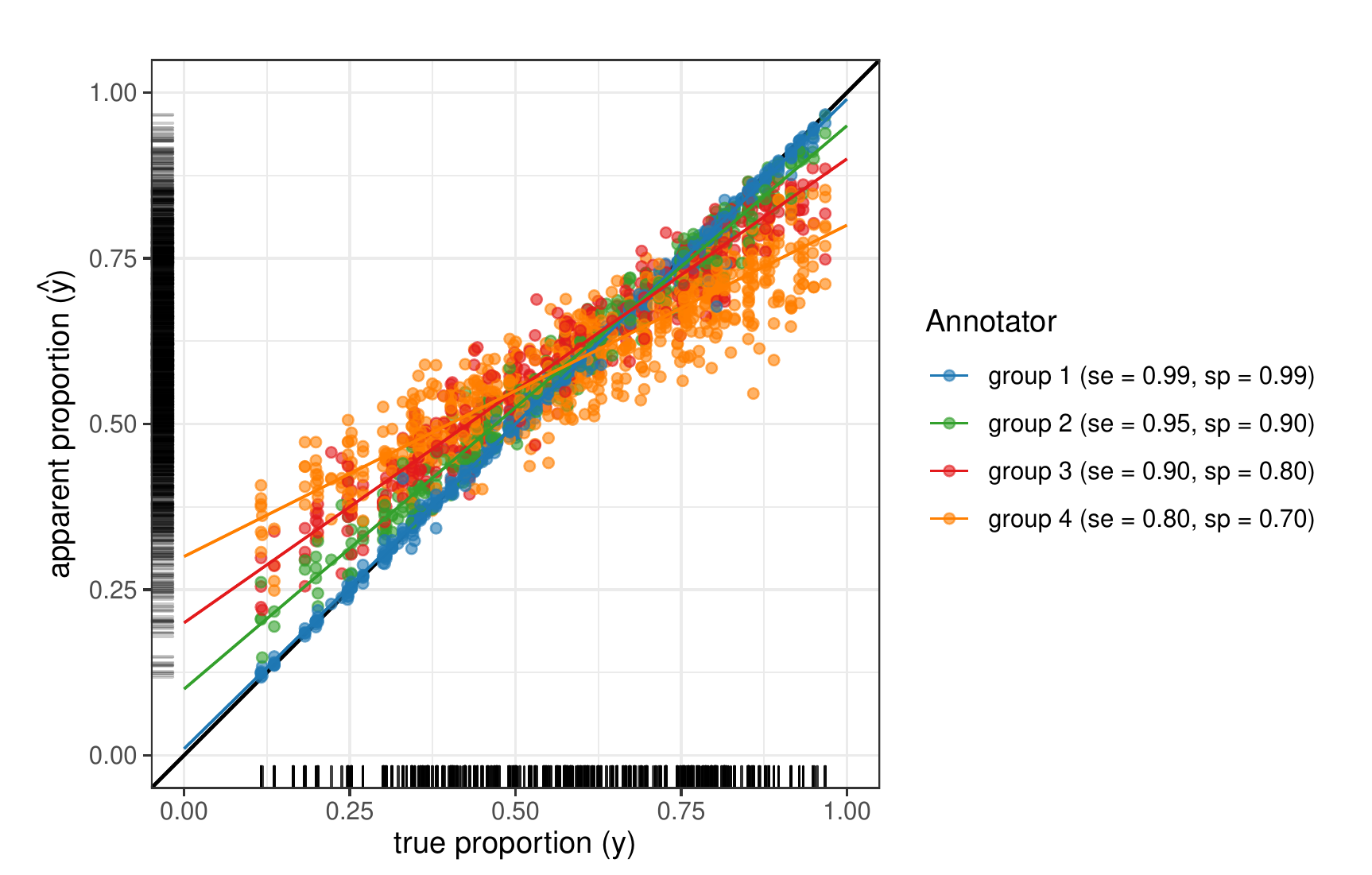} 
		\vspace{-0.5cm}
                \caption{ Apparent proportion ($\hat{y}_{ij}$) in four groups of users vs the true ($y_j$)
                  proportion. 
The black diagonal line represents the ideal case of perfect classification. The lines give the $\hat{y}_{ij}$ for each value of $y_j$ based on the users' mean $se_i$ and $sp_i$, with points representing the simulated values. 
		Subjects in group 1 have mean $se = 0.99$, $sp = 0.99$; subjects 
		in 2 have $se = 0.95$, $sp = 0.90$; subjects in 3 have
        $se = 0.90$, $sp = 0.80$; and subjects in 4 have
        $se = 0.80$, $sp = 0.70$. In all of groups the precision $\phi_{se} = 50$ and $\phi_{sp} = 50$.}
\label{fig:4Fig4}
\end{figure}

For a better understanding of the influence of the performance
measures, we show $\hat{y}_{ij}$ as a function of the true latent
variable $y_j$ for the four groups of subjects in Fig.~\ref{fig:4Fig4}. The
diagonal black solid line represents the ideal case of perfect
classification ($se_i = sp_i = 1$).  For the subjects with a beginner's
skill level (group 4),
$se_i = 0.8$ and $sp_i = 0.7$ and as a result, the apparent proportion is substantially
different to the true value, mostly for small and large values of
$y_j$. For example, if $y_j = 0.10$, the elicitation of apparent
proportion $\hat{y}_{ij}$ was more than three times this value (0.35).
Overestimation ($\hat{y}_{ij} > y_j$) occurs when
$p < \left ( 1-sp_i \right )/\left ( 2-se_i-sp_i \right)$, and
underestimation ($\hat{y}_{ij} < y_j$) otherwise. When $se_i = sp_i$,
$\hat{y}_{ij} > y_j$ if $y_j < 0.5$.

\subsection{Bayesian data analysis}

The aim of the Bayesian data analysis is to learn about $y_j$ and the
regression coefficients ($b_0$ and $b_1$) for the two models (weighted
and SDME). Additionally, we want to estimate the subject's
classification performance measures ($se_i$ and $sp_i$) in the SDME
model. We used Hamiltonian Monte Carlo (HMC) simulations in Stan
\citep{carpenter2017stan}, which is based on the no-U-turn sampler
(NUTS) \citep{hoffman2014no}. We used 3 chains each with 60,000
samples, discarded a burn-in of 30,000 samples, and used a thinning
rate of 1 in 3.

The parameters $se_i$ and $sp_i$ in the SDME model and $acc_i$ in the weighted model can only be weakly identified, since the model involves more parameters than can be well estimated from the dataset. 
Thus informative priors obtained from the testing dataset (where $y_j$ is
known) were used for these parameters. 

A sum-to-zero constraint was also imposed on the spatial component
$u_j$ for $j = 1, 2, \cdots, m$. Normal priors were used for the regression coefficients
with the mean
obtained from the maximum likelihood beta regression in R \citep{betareg}.  Finally, a
truncated normal prior was set for the precision parameter in the beta
distribution ($\phi \sim \mathcal{N}(20, 5)T[10, 60])$. 
Details of the model are given in Fig.\ref{fig:priors}.
The Stan code for the SDME model is provided in the supporting materials.

\begin{figure}[h]
	\centering
	{\scriptsize
\begin{align*}
\hat{y}_{ij} & = y_{j} \times se_{i} + (1 - y_{j})\times(1 - sp_{i}) && \\	
y_j | \alpha_j, \beta_j  &\sim \textrm{Beta}(\alpha_j,\beta_j ) && \\
\alpha_j & = \mu_j \phi  && \\
\beta_j & = -\mu_j \phi + \phi  && \\
\textrm{logit}(\mu_j) &= X_jb + u_j && \\ 
u_{l}|u_{t},\tau_u &\sim \mathcal N \left ( \frac{1}{n_l}\sum_{l \sim t}^{.}u_t, \frac{1}{\tau_u n_l} \right )&& \\
\mu_{j} & = \E[y_j|\alpha_j, \beta_j]  && \\
\V(y_j) & = \frac{\mu_{j}(1-\mu_{j})}{(1+\phi)}  && \\
\textrm{Priors}&  && \\
    b_{0} & \sim \mathcal{N}\left(\hat{\mu}_{b_0},5\right)  && \text{\# informative prior on the regression coefficient intercept}\\	
		b_{1} & \sim \mathcal{N}\left(\hat{\mu}_{b_1},5\right)  && \text{\# informative prior on the regression coefficient slope}\\	
		se_{i}&\sim \textrm{Beta}\left(\alpha_{se},\beta_{se}\right)   && \text{\# hierarchical informative prior on the sensitivities}\\
		sp_{i}&\sim \textrm{Beta}\left(\alpha_{sp},\beta_{sp} \right)  && \text{\# hierarchical informative prior on the specificities}\\
		\phi &\sim \mathcal{N}\left(20,5\right)T[10,60]  && \text{\# prior on the beta distribution precision}\\	
		u_{j} &\sim \textrm{CAR}\left(\tau_u, W, D\right)  && \text{\# CAR prior for the spatial model}\\		
		\tau_{u}  & \sim \textrm{Gamma}\left(0.1, 0.1\right) && \text{\# precision of the spatial effect (CAR prior) } \\	
		\label{eq:66}
\end{align*} 
}%
\vspace{-1.9cm}
	\caption{Hierarchical SDME model and prior distributions.  }%
	\label{fig:priors}
\end{figure}

\begin{figure}[h]
	\centering
	{\scriptsize
\begin{align*}
y_{ij} | \alpha_{ij}, \beta_{ij}  &\sim \textrm{Beta}(\alpha_{ij},\beta_{ij} ) && \\
\alpha_{ij} & = \mu_j w_{i} \phi  && \\
\beta_{ij} & = -\mu_j w_{i} \phi + \phi  && \\
\textrm{logit}(\mu_j) &= X_jb + u_j + \varepsilon_j && \\
u_{l}|u_{t},\tau_u &\sim \mathcal N \left ( \frac{1}{n_l}\sum_{l \sim t}^{.}u_t, \frac{1}{\tau_u n_l} \right )&& \\
w_{i} & = 1 / acc_{i}  && \\
\mu_{j} & = \E[y_j|\alpha_j, \beta_j]  && \\
\V(y_j) & = \frac{\mu_{j}(1-\mu_{j})}{(1+\phi)}  && \\
\textrm{Priors}&  && \\
    b_{0} & \sim \mathcal{N}\left(\mu_{b_0},5\right)  && \text{\# informative prior on the regression coefficient intercept}\\	
		b_{1} & \sim \mathcal{N}\left(\mu_{b_1},5\right)  && \text{\# informative prior on the regression coefficient slope}\\	
		acc_{i}&\sim \textrm{Beta}\left(\alpha_{se},\beta_{se}\right)   && \text{\# hierarchical informative prior on the accuracy}\\
				\phi &\sim \mathcal{N}\left(20,5\right)T[10,60]  && \text{\# prior on the beta distribution precision}\\	
		u_{j} &\sim \textrm{CAR}\left(\tau_u, W, D\right)  && \text{\# CAR prior for the spatial model}\\		
		\tau_{u}  & \sim \textrm{Gamma}\left(0.1, 0.1\right) && \text{\# precision of the spatial effect (CAR prior) } \\	
\end{align*} 
}%
\label{eq:777}
\vspace{-1.9cm}
	\caption{Hierarchical weighted model and prior distributions.  }%
	\label{fig:priors}
\end{figure}

\subsection{Simulation results}
\label{sec:res}

The SDME model captured the true parameter values much better than the weighted model (Table \ref{table:tabpost}).  
The estimates for $\phi$ are off in the weighted model and the SE of the mean and 95\% density intervals are narrow. 
The intervals are wider for SDME, but they capture the true parameters' values.

The posterior densities and trace plots for the regression
coefficients show well mixed chains and apparent convergence
(See the supporting materials) and 
the R-hat convergence diagnostic \citep{gelman1992inference, vehtari2019rank} produced values well below 1.1. 
The posterior values of $se_i$ and $sp_i$ values were equally well retrieved for most of the
subject groups (also in supporting materials). 

\begin{table}
\caption{\label{table:tabpost} The true value for each of the model parameters, as well as the summary statistics for their posterior distribution, including the mean, standard error (se\_mean), and percentiles. The column se\_mean represents the Monte Carlo standard error.}
\scalebox{0.80}{
\begin{tabular}{rrrrrrrrrrr} 
  \hline
model & param& true value & mean & SE mean & sd & 2.5\% & 25\% & 50\% & 75\% & 97.5\% \\ 
  \hline
	
 weighted & $b_{0_w}$ & 1 & 0.909 & 0.010 & 0.052 & 0.813 & 0.874 & 0.907 & 0.942 & 1.023 \\ 
  weighted & $b_{1_w}$ & -2& -1.488 & 0.029 & 0.154 & -1.836 & -1.573 & -1.486 & -1.377 & -1.217 \\ 
  weighted & $\phi$ & 30 & 59.909 & 0.003 & 0.091 & 59.677 & 59.873 & 59.937 & 59.974 & 59.998 \\ 
  weighted & $acc_1$ &0.991 & 0.993 & 0.000 & 0.009 & 0.968 & 0.990 & 0.996 & 0.999 & 1.000 \\ 
  weighted & $acc_2$ &0.924 & 0.916 & 0.001 & 0.027 & 0.860 & 0.899 & 0.918 & 0.935 & 0.962 \\ \hline
 
 	SDME & $b_{0_{SDME}}$ &1 & 1.023 & 0.002 & 0.091 & 0.847 & 0.961 & 1.021 & 1.083 & 1.204 \\ 
  SDME & $b_{1_{SDME}}$ &-2 & -2.020 & 0.007 & 0.230 & -2.473 & -2.171 & -2.020 & -1.867 & -1.568 \\ 
  SDME & $\phi$ & 30 & 28.009 & 0.098 & 4.450 & 19.728 & 24.924 & 27.834 & 30.917 & 37.118 \\ 
  SDME & $se_1$ & 0.990 &0.991 & 0.000 & 0.010 & 0.963 & 0.987 & 0.995 & 0.998 & 1.000 \\ 
  SDME & $se_2$ & 0.948&0.948 & 0.000 & 0.031 & 0.873 & 0.931 & 0.954 & 0.971 & 0.990 \\ 
	SDME & $sp_1$ & 0.991&0.992 & 0.000 & 0.011 & 0.962 & 0.988 & 0.995 & 0.999 & 1.000 \\ 
  SDME & $sp_2$ & 0.899&0.898 & 0.001 & 0.039 & 0.809 & 0.875 & 0.903 & 0.927 & 0.961 \\ 
  \hline

\end{tabular}
}
\end{table}

\begin{landscape}

\begin{figure}[hptb]
	\centering
	\vspace{-0.5cm}
		\includegraphics[width=10.5in]{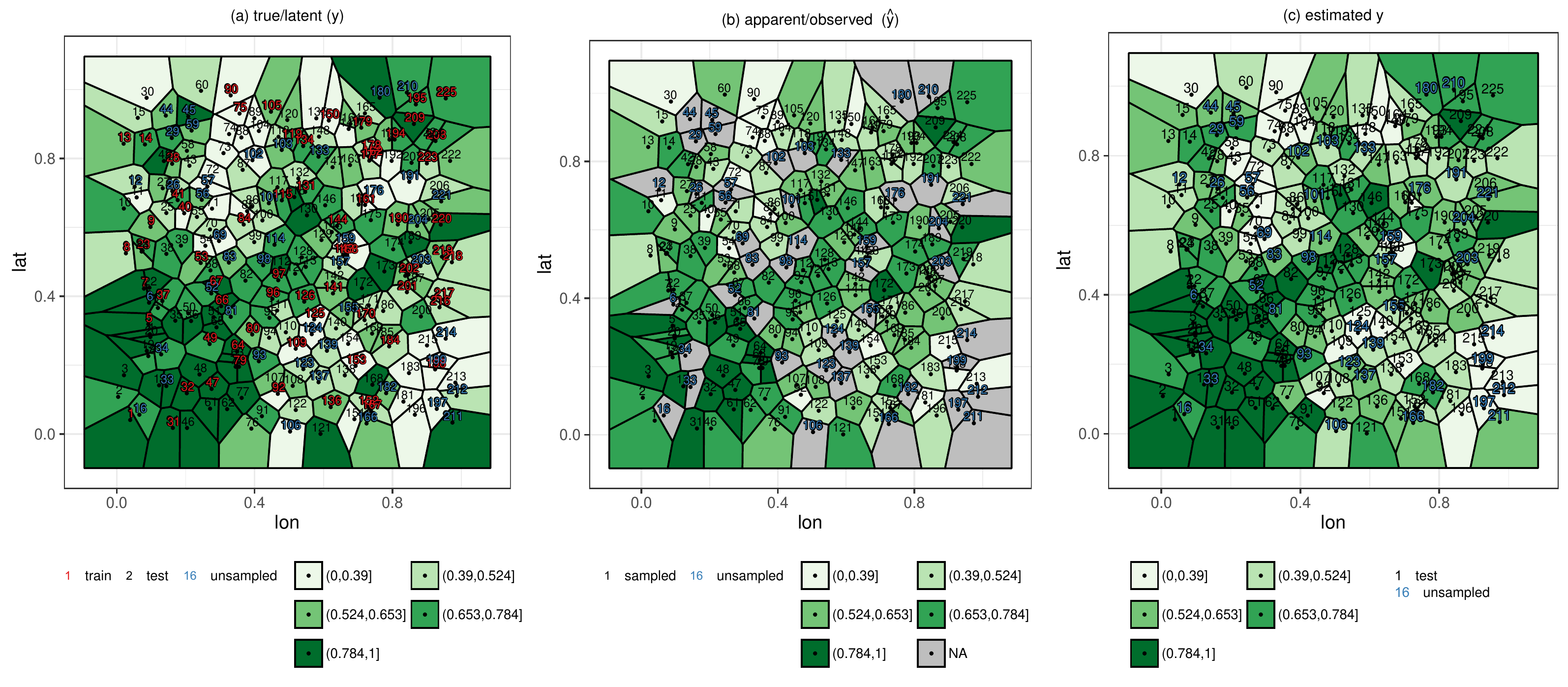}
                \caption{Voronoi diagram of (a) the true latent
                  proportion $y_j$ at 225 simulated locations,
                  (b) the apparent proportion elicited by subjects
                  ($\hat{y}_{ij}$), and (c) estimated latent fraction ($y_{\textrm{estim}_j}$). 
									The exact class of the category of $y_j$ is
                  obtained in 199 out of 225 locations. 
                  The blue numbers are the cells where values of $\hat{y}_{ij}$ are missing and
$y_j$ was predicted solely using the covariate $x$ and neighbouring
information.
}
	\label{fig:15Fig15}
\end{figure}
\end{landscape}

Fig.~\ref{fig:15Fig15} (c) depicts the degree of success in retrieving the true latent proportion ($y_j$). The diagram in (c) contains the estimated values of $y_j$, which resembles the true latent pattern from (a).  
The posterior distributions of $y_j$ in 220 cells out of 225 (97.78\%) fall within the 95\% highest density interval, which shows a suitable coverage of these estimates.
Equally the true latent class is well retrieved in 31 of out of 45 unsampled locations, 
(text in blue color) comparing (c) with (a).

\begin{figure}[hp]
	\centering
		\includegraphics[width=4.5in]{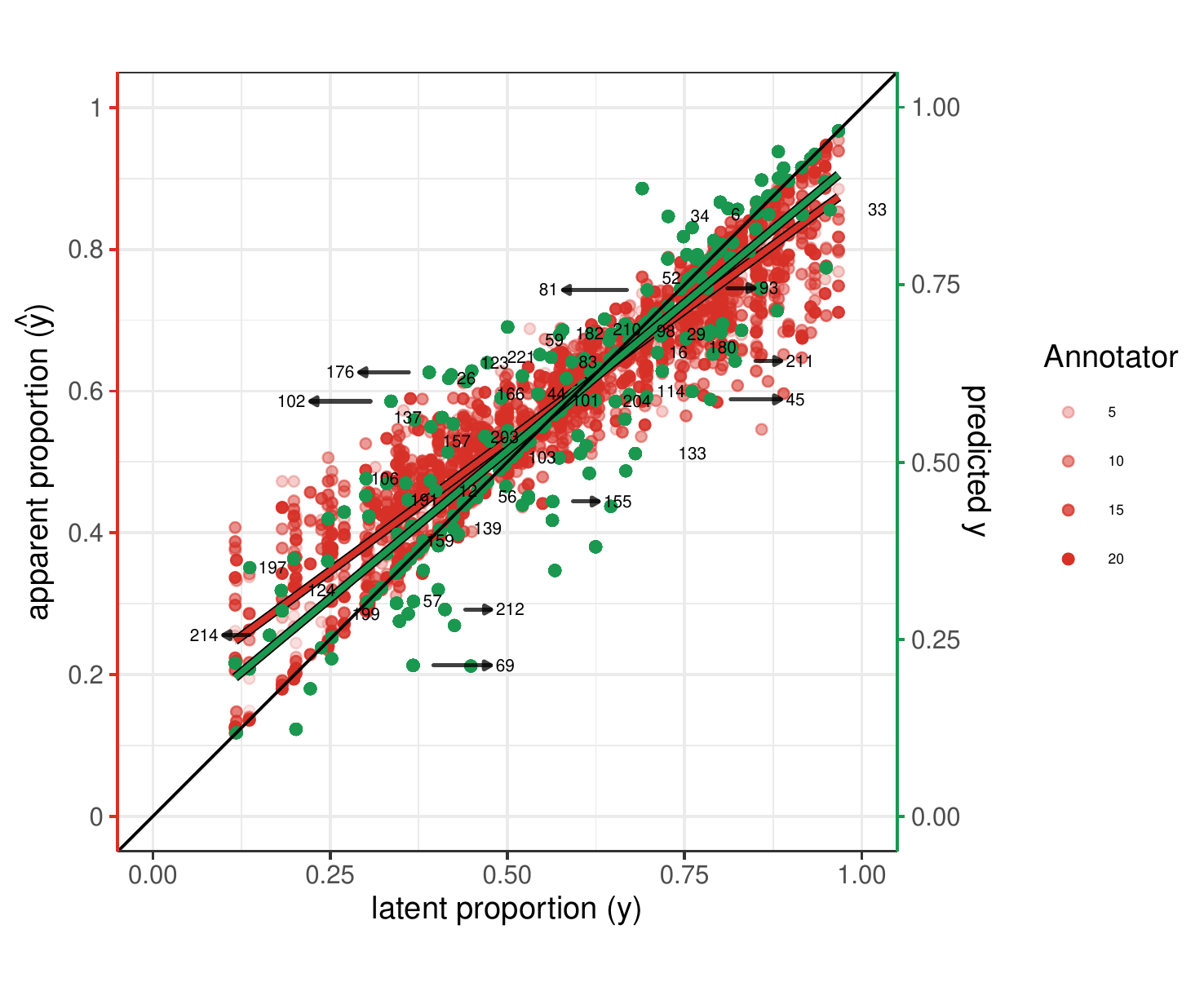}
		\vspace{-0.5cm}
\caption{Bias correction obtained from the spatially
dependent misclassification error (SDME) model. 
The red dots are the elicited $\hat{y}_{ij}$ as a function of the true $y_{j}$ in the four groups of users (dark red = low $se_i$ and $sp_i$). 
The green dots shows the estimated latent variable ($y_{\textrm{estim}}$) vs
$y_j$ after accounting for misclassification errors.
The image id of the prediction in 45 unsampled locations
from Fig.~\ref{fig:1Fig1} (b) is also shown. 
The red and green solid lines are the regression lines from the apparent proportion $\hat{y}_{ij}$ and estimated $y_{\textrm{estim}_j}$ from the model showing how the bias is corrected.
The arrows identify the points in the unsampled dataset.}
	\label{fig:12Fig12}
\end{figure}

Accounting for misclassification errors allows us to 
correct the bias in $\hat{y}_{ij}$ and obtain suitable estimates of $y_j$ (Fig.~\ref{fig:12Fig12}). In this figure the red dots are the original $\hat{y}_{ij}$ values obtained from the subjects' classifications.
The green dots are the estimated values of $y_j$ when we account for misclassification errors.

A good identification is achieved, with a large proportion of the green dots falling
along the diagonal solid line showing a suitable precision of the estimates. The image identifiers for the 45 unsampled locations from Fig.~\ref{fig:1Fig1}
(b) are shown in black.  In these points $y_j$ is estimated from the
neighbouring measurements and the covariates. 
There are a few points where the difference in the $y_j$ and $\hat{y}_{ij}$ values is relatively large
due to spatial dependency causing the predicted proportion to be higher or lower than expected (e.g. Fig.~\ref{fig:12Fig12}, point 69).

\newgeometry{inner=1.8cm,outer=1.8cm, top=1.8cm, bottom=1.8cm}
\restoregeometry

\begin{figure}[h]
	\centering
		\includegraphics[width=4in]{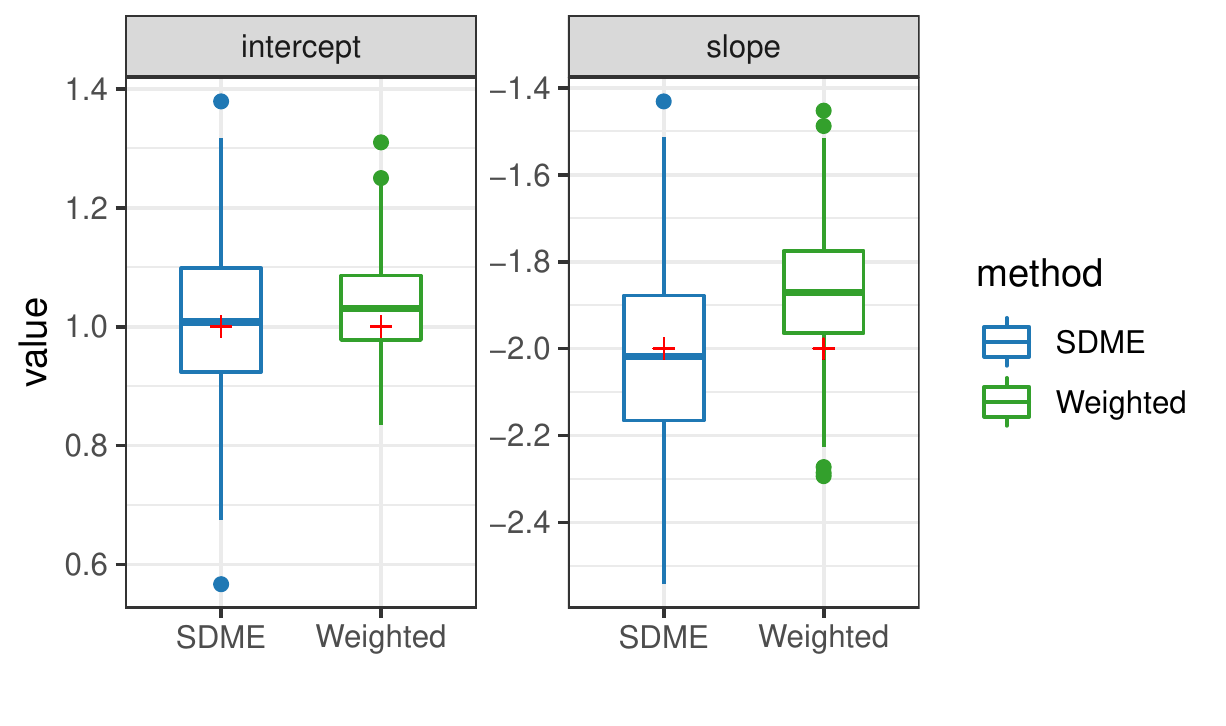}
                \caption{Boxplots and violin plots of the posterior regression
                  coefficient estimates from weighted and SDME
                  regression models fit to 300 random simulated
                  datasets. The red dots indicate the true parameter values.}
\label{fig:13Fig13}
\end{figure}

\begin{figure}[h]
	\centering
		\includegraphics[width=6in]{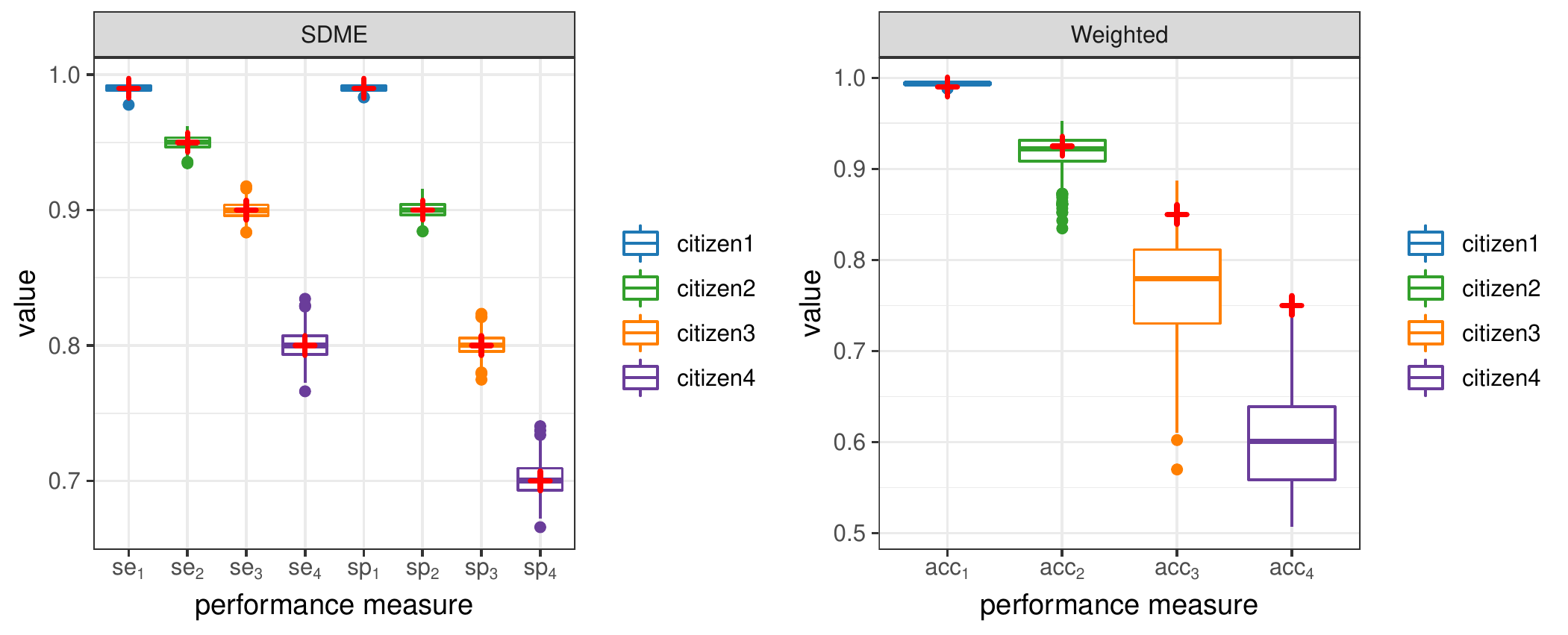}
	\caption{Boxplots of the posterior performance measures in the SDME ($se_i$ and $sp_i$) and weighted ($acc_i$) models for the four groups of subjects in the 300 SDME models. The red crosses indicate the true values.} 
\label{fig:14Fig14}
\end{figure}

We fitted the weighted and SDME models to each of the 300 simulated datasets and compared how well they retrieved the parameters of interest.
The posterior means for the beta regression coefficients suggest that overall
the SDME model produced suitable estimates for the
regression parameters (Fig.~\ref{fig:13Fig13}). 
As expected the weighted regression produced more biased regression coefficients than the SDME model, especially for the slope.
The SDME model also produced suitable estimates of the subjects' $se_i$ and $sp_i$ values (Fig.~\ref{fig:14Fig14}). 
However, poor performance measure estimates are obtained from the weighted model.

\clearpage
\section{Case study results} %
\label{sec:cs} %

\begin{figure}[htbp]
	\centering
		\includegraphics[width=6.0in]{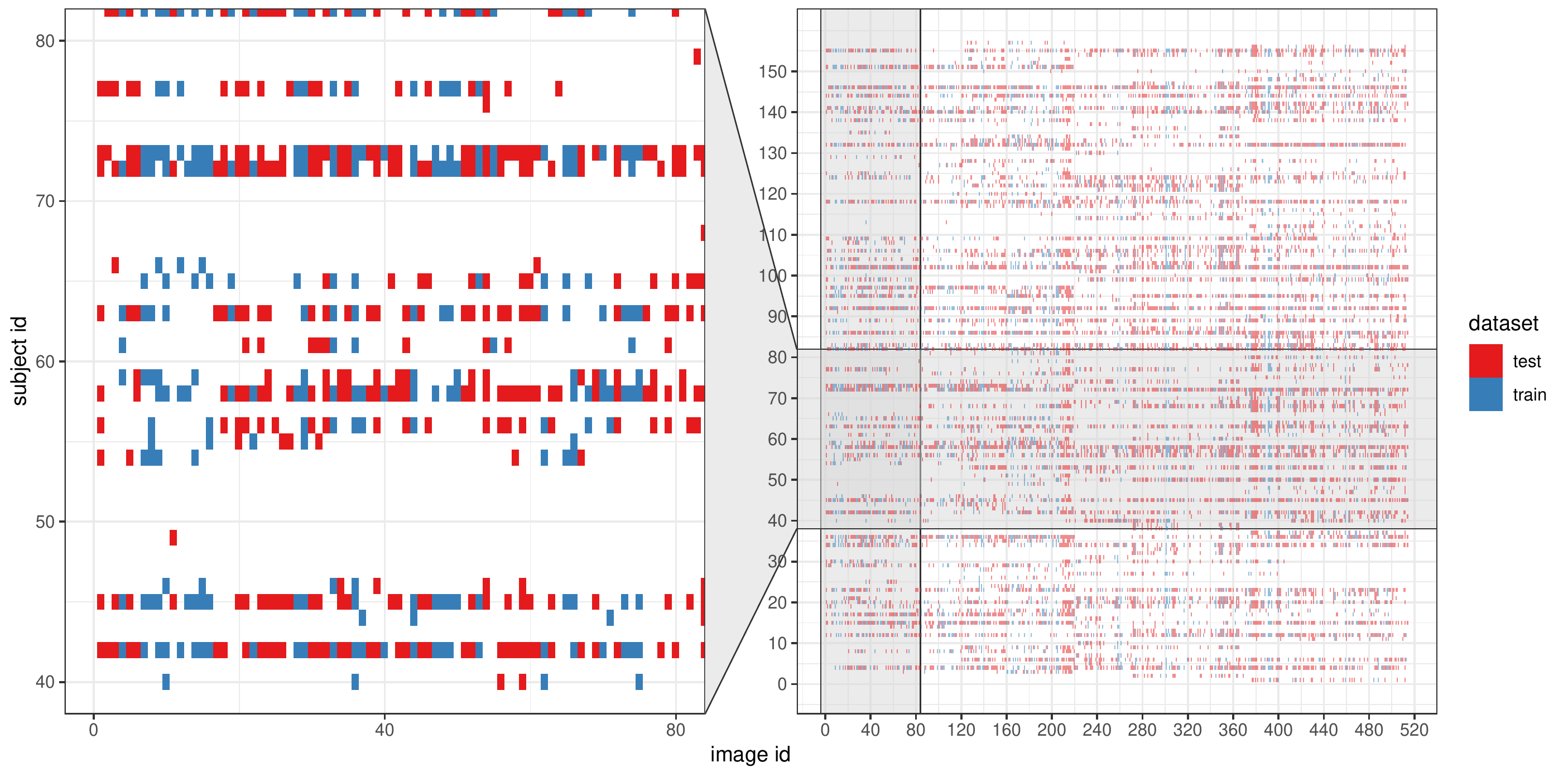}
                \caption{Images classified per user. The plot on the
                  left shows a subset of the whole dataset, which
                  is shown on the right. Images in red were used
                  as training to learn about the users' performance
                  measures. In the blue images, the underlying true
                  value is unknown and predicted by the model
                  (testing).}
\label{fig:21Fig21usersmedia}
\end{figure}

We applied the SDME model to coral cover data elicited from underwater images from the GBR.
We used the training dataset composed of 171 images to obtain estimates of each subject's $se_i$ and $sp_i$ mean values, which were used to obtain informative priors for the hierarchical model. 
For example, from Fig.~\ref{fig:21Fig21usersmedia} subject 40 classified images 10, 36, 56, 59, 62, 75, etc. 
Using the classification from the training images (10, 36, 62, 75, etc.) we obtained beta prior distributions for $se_{i = 40}$ and $sp_{i = 40}$, which were then used to predict the false positive/negative rates. 
This allowed us to estimate the unknown $y_j$ in the test and unsampled
datasets and produce posterior distributions for the $se_i$ and $sp_i$
parameters.
In these images, information was also borrowed from the neighbours based on the CAR model and from the four covariates introduced in Section \ref{sec:motiv}
Weakly informative priors were also obtained for the regression coefficients from fitting a non-Bayesian beta regression to the data.   

The model was fit using Stan on a HPC node with 65 Gb of memory using three processors.
We used three chains, a warm-up period of 12,000 samples out of 24,000 iterations and preserving one in three samples (thinning).

\begin{figure}[hp] 
	\centering
		\includegraphics[width=4.5in]{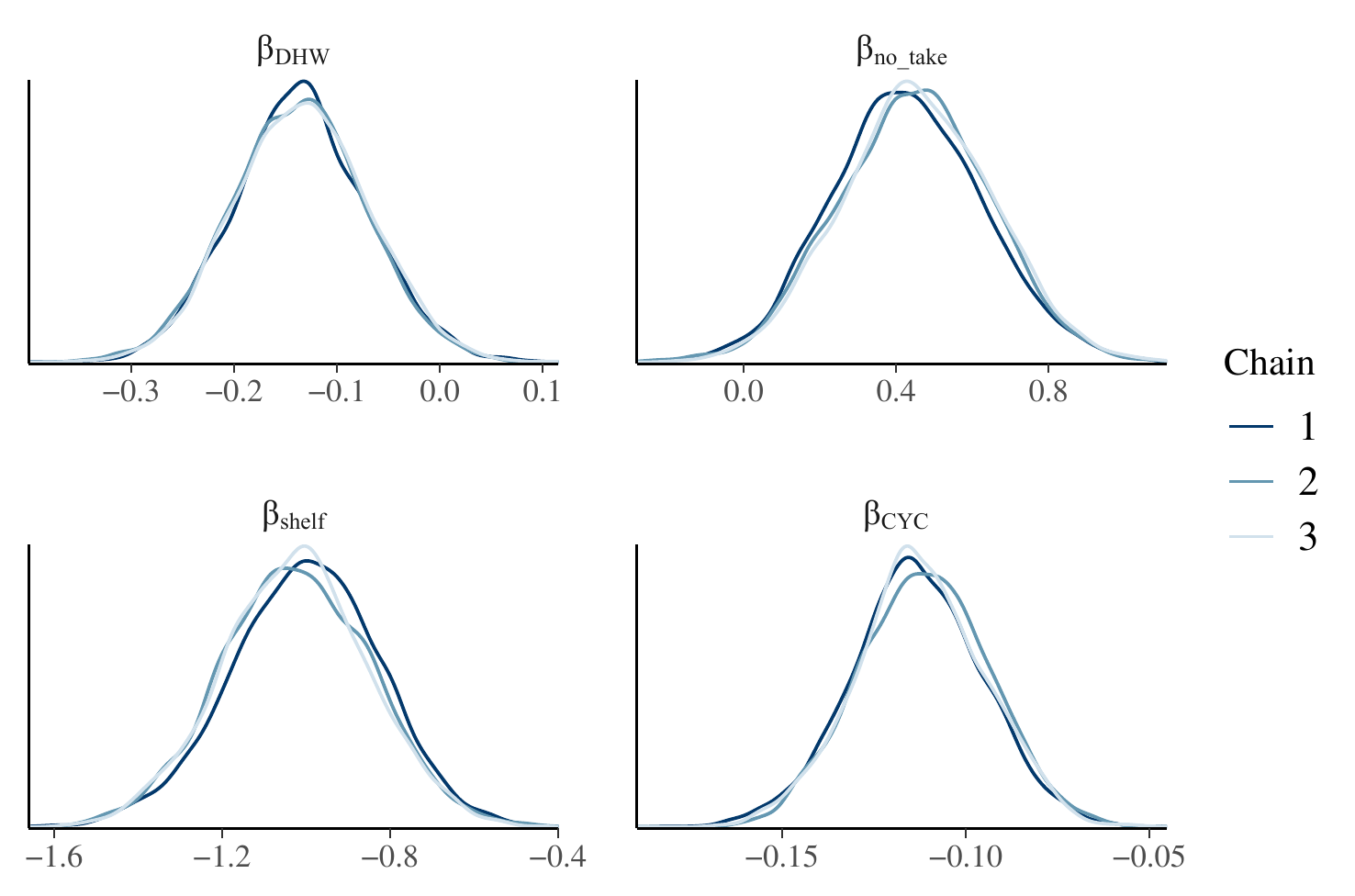}
	\caption{Posterior densities of the regression coefficients.} 
\label{fig:23Fig23}
\end{figure}

\begin{figure}[htbp]
	\centering
		\includegraphics[width=4in]{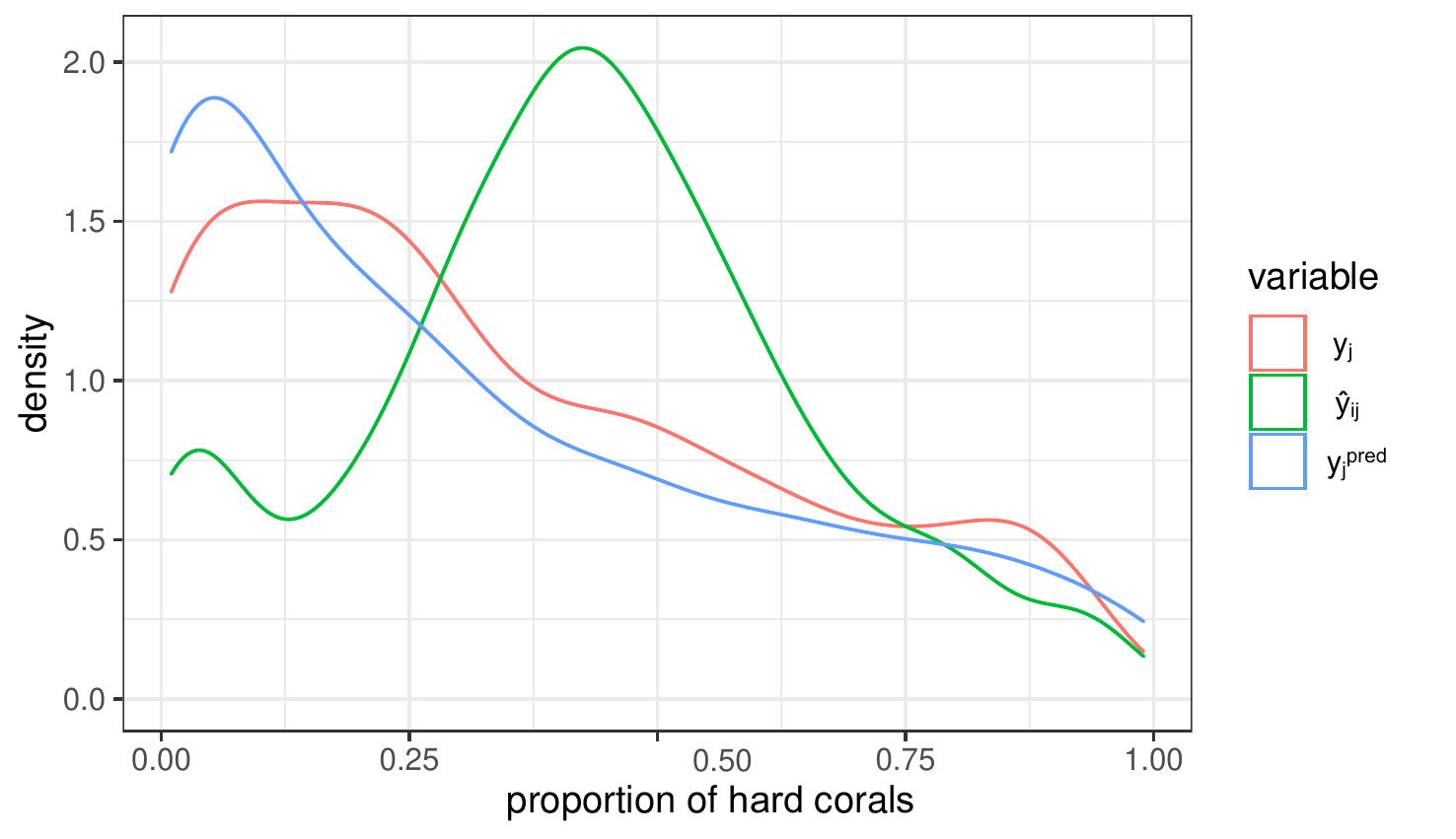}
    \caption{Density of the true latent proportion of hard corals $y_j$ (in red). The apparent proportion $\hat{y}_{ij}$ and 
the posterior mean of the predicted proportion $y_{j_{pred}}$ are shown in green and blue, respectively.}
\label{fig:20Fig20success}
\end{figure}

Table \ref{table:tabpost2} shows the summary statistics of the posterior distributions for the parameters of interest. 
The DHW and cyclone impact covariates have a substantial negative effect on the proportion of hard corals, with the 97.5\% credible interval well below 0. 
On the other hand, no-take marine reserves and middle shelf reefs tend to have substantially higher proportions of corals. The posterior densities of the regression coefficients are shown in Fig.~\ref{fig:23Fig23}.
These findings are in agreement with several other studies \citep[e.g.][]{hughes2018global, beeden2015impacts}  
and demonstrate the capacity of crowdsourced data to answer relevant ecological questions. 
Fig.~\ref{fig:20Fig20success} shows the latent and the apparent densities $y_j$ and $\hat{y}_{ij}$ respectively and the estimated posterior density of $y_j$. 
It shows how well the model corrects the bias 
in the apparent proportion $\hat{y}_{ij}$,
producing estimates $y^{\textrm{pred}}_j$ quite similar to the latent proportion $y_{j}$.

\begin{table}
\caption{\label{table:tabpost2} Summary statistics of the posterior distribution of the model parameters
 including the mean, Monte Carlo standard error, standard deviation (sd), and the percentiles.
Estimates of the latent variable ($y_j$) are also given for the first four locations.
}
	\scalebox{0.8}{
\centering
\begin{tabular}{rrrrrrrrr} 
  \hline
	parameter & mean & se\_mean & sd & 2.5\% & 25\% & 50\% & 75\% & 97.5\% \\    \hline
$b_{DHW}$ & -0.136 & 0.001 & 0.065 & -0.262 & -0.179 & -0.136 & -0.092 & -0.007 \\ 
$b_{notake}$ & 0.444 & 0.004 & 0.199 & 0.061 & 0.310 & 0.443 & 0.579 & 0.836 \\ 
$b_{shelf}$ & -1.017 & 0.004 & 0.176 & -1.370 & -1.136 & -1.015 & -0.896 & -0.674 \\ 
$b_{cyclone}$ & -0.112 & 0.000 & 0.018 & -0.147 & -0.124 & -0.113 & -0.100 & -0.078 \\ 
 $\phi$ & 10.897 & 0.100 & 1.632 & 10.017 & 10.190 & 10.474 & 11.038 & 13.998 \\ 
  $se_1$ & 0.838 & 0.001 & 0.134 & 0.540 & 0.748 & 0.873 & 0.953 & 0.998 \\ 
  $se_2$ & 0.872 & 0.001 & 0.136 & 0.543 & 0.789 & 0.925 & 0.986 & 1.000 \\ 
  $se_3$ & 0.869 & 0.001 & 0.135 & 0.546 & 0.787 & 0.918 & 0.982 & 1.000 \\ 
  $se_4$ & 0.855 & 0.001 & 0.134 & 0.541 & 0.771 & 0.896 & 0.969 & 0.999 \\ 
  $sp_1$ & 0.902 & 0.001 & 0.126 & 0.567 & 0.847 & 0.963 & 0.997 & 1.000 \\ 
  $sp_2$ & 0.876 & 0.001 & 0.128 & 0.555 & 0.804 & 0.922 & 0.982 & 1.000 \\ 
  $sp_3$ & 0.815 & 0.001 & 0.136 & 0.529 & 0.714 & 0.841 & 0.932 & 0.995 \\ 
  $sp_4$ & 0.787 & 0.001 & 0.134 & 0.526 & 0.682 & 0.801 & 0.901 & 0.987 \\ 
  $y_1$ & 0.346 & 0.003 & 0.266 & 0.002 & 0.116 & 0.293 & 0.535 & 0.913 \\ 
  $y_2$ & 0.310 & 0.003 & 0.240 & 0.003 & 0.107 & 0.262 & 0.470 & 0.853 \\ 
  $y_3$ & 0.382 & 0.003 & 0.267 & 0.006 & 0.154 & 0.342 & 0.583 & 0.927 \\ 
  $y_4$ & 0.374 & 0.003 & 0.272 & 0.003 & 0.139 & 0.328 & 0.581 & 0.933 \\ 
   \hline

\end{tabular}
}
\end{table}

\section{Discussion and conclusions}
\label{sec:Dis}

Modern ecological research is relying more and more on citizen science
data to learn about latent variables such as the prevalence and
abundance of key species and communities \citep{delaney2008marine, van2013opportunistic, bird2014statistical, bain2016citizen}. 
However, data elicited from citizens is likely to be imprecise and biased \citep{isaac2014statistics, burgess2017science}. 
We present a spatial Bayesian hierarchical model to account for misclassification errors, as well as spatial dependence in the data. The SDME approach  
can be applied in many other ecological and conservation studies, where citizen scientists are asked to classify images, videos, audio files, etc. For example, in the classification of species presents in videos, the identification birds from audio recordings, etc. 

The SDME approach provides a number of benefits over a weighted modelling approach (Table \ref{table:compara}). 
It produces more precise regression coefficient estimates, allows the estimation of the latent variable of interest and accounts for the subjects' abilities.  
In addition, the case study corroborated results from previous studies, which were obtained using data from professional monitoring programs \citep{de201227, hughes2018global, beeden2015impacts}. Although our results did not reveal new ecological relationships, they do suggest that the SDME model can be used to gain a more accurate understanding of the relationships between ecologically meaningful covariates and the response. Despite these advantages, there are also limitations to the SDME approach (Table \ref{table:compara}). 

First, the model is slightly more computationally intensive
and becomes prohibitive when the number of locations and elicitations
is large (i.e. scalability issues). 
Second, the SDME model contains a larger number of parameters and as a result, the model cannot be properly identified unless (weakly) informative priors are used for some of them. 
Finally, both models rely on a large number of classifications per image to ensure that the estimates are robust. 
 
\begin{table}
\caption{\label{table:compara} Comparison of the weighted regression vs the spatially dependent misclassification error (SDME) model.}
\centering
\scalebox{0.890}{
\begin{tabular}{lll} 
\hline
Model             & benefits 																	& disadvantages\\ \hline
Weighted model    & slightly faster					 												& biased regression coefficients\\ 
				& less parameters to estimate 							& no estimates of users' abilities or latent variable\\ 
				&       											 		& more suitable when users have nearly perfect abilities\\ \hline
SDME model  	&  more precise regression coefficient estimates	& more identifiability issues\\
				&  accounts for and estimates the users' abilities 			
				& more computationally expensive\\
				&  produces estimates of the latent response variable &  \\
\hline

\end{tabular}
}

\end{table}

The SDME approach presented here outperformed the weighted approach, but there are other modelling approaches that could also be used. For example, the SDME approach can also be expressed in terms of points containing hard corals rather than a proportion, which can be modelled using discrete distributions such as binomial, Poisson or negative binomial.
Similarly, other alternatives for the spatial random effect could be used such as Gaussian random fields or Gaussian processes based on nearest neighbours e.g. \citet{datta2016hierarchical, finley2017applying}, etc. could be considered.  We also explored other approaches for estimating the latent response ($y_j$). For example, the labels at the point level can be obtained from the majority vote; however, this does not work well for difficult tasks such as those presented in our case study because the probability of answering correctly could be low. Another possible formulation results from rearranging  Eq \ref{eq:yhat} where $y_j$ is directly estimated from the apparent proportion $\hat{y}_{ij}$, $se_i$ and $sp_i$. This approach also results in suitable parameter estimates, but requires rather precise $se_i$ and $sp_i$ distributions.

Multiple extensions to the model here discussed can be implemented. 
For example, we can consider the user's $se_i$ and $sp_i$ distributions to be a mixture, 
which would be affected by the underlying task difficulty. 
This is in line with what \citet{chambert2018two} proposed. 
The false-positive rate tends to be higher when closely related categories are present. 
Similarly spatio-temporal extensions can be implemented as more images across years become available. 
In our case study, images were obtained from a professional monitoring program. 
However, geographical or spatial recording bias should be considered when the data are collected opportunistically \citep{van2013opportunistic, isaac2014statistics, mair2017evaluating}.
Further investigations are also needed to assess whether the results are robust to the way 
we define the areal units.
Another variation could include recursive Bayesian estimation
\citep{sarkka2013bayesian} in which the model is updated as new data
become available.  
Finally, the integration of coral cover data from CS programs and professional and scientific
monitoring programs could also strengthen the model and produce more
precise and robust estimates.

\section*{R/Stan codes and data used in the study}
The R/Stan codes are hosted in \url{https://github.com/EdgarSantos-Fernandez/reef_misclassification}.
The data used in the case study can be obtained on request from the first author.

\section*{Acknowledgement}
Thank you to the four reviewers and the editor for their very insightful, detailed and constructive comments, which allowed us to substantially improve the manuscript.
The authors declare that they have no conflicts of interest. 
This research was supported by the
Australian Research Council (ARC) Laureate Fellowship Program under the project ``Bayesian Learning for Decision Making in the Big Data Era'' (ID: FL150100150) and
the Centre of Excellence for Mathematical and Statistical Frontiers (ACEMS). 
Thanks to the members of the VRD team (\url{https://www.virtualreef.org.au/about/}) in particular to Bryce Christensen.
We also thank all the workers that contributed to the classification of images.
Ethics approval was obtained from the QUT Human Ethics Advisory Team (Number 1600000830).
Some of the computations were performed through the QUT High Performance Computing (HPC) infrastructure.
Data analysis and computations were undertaken in \textsf{R} software \citep{rprog} using the packages \textsf{rstan} \citep{rstan}. 
Data visualizations were made with the packages \textsf{tidyverse} \citep{tidyverse}, \textsf{bayesplot} \citep{bayesplot}, \textsf{ggvoronoi} \citep{ggvoronoi}, 
\textsf{ggrepel} \citep{ggrepel}, and \textsf{ggforce} \citep{ggforce}.


\section*{Appendix}
\subsection*{Glossary of symbols and definitions}
\label{sec:symb}

\begin{table}
\caption{\label{table:abb} Symbols and definitions}
\scalebox{0.80}{
\begin{tabular}{ll} 
\hline
$i$ & subject id, $1,2,\cdots, n$\\
$j$ & image number, $1,2,\cdots, m$\\
$k$ & elicitation point, $1,2,\cdots, q$\\

$z_{ijk} = \{0,1\}$ & indicates if the $k^{th}$ point in the image $j$ was classified as hard coral by the subject $i$ \\
$q$ & number of elicited points in the image $j$\\

$y_j | \alpha_j, \beta_j  \sim \textrm{Beta}(\alpha_j,\beta_j )$	&	true coral cover (latent variable) of the image $j$\\
$\alpha_j = \mu_j \phi$	&	first shape parameter in the beta distribution on the image $j$\\
$\beta_j = -\mu_j \phi + \phi$	&	second shape parameter in the beta distribution on the image $j$\\

$\hat{y}_{ij} = y_{j} \times se_{i} + (1 - y_{j})\times(1 - sp_{i})$	&	apparent coral cover (elicited variable) on the image $j$ by the subject $i$  \\

$se_i = \frac{\sum_{j=1}^{m}\sum_{k=1}^{q}TP_{ijk}}{\sum_{j=1}^{m}\sum_{k=1}^{q}TP_{ijk} + \sum_{j=1}^{m}\sum_{k=1}^{q}FN_{ijk}}$ & sensitivity from the user $i$\\
$sp_i = \frac{\sum_{j=1}^{m}\sum_{k=1}^{q}TN_{ijk}}{\sum_{j=1}^{m}\sum_{k=1}^{q}TN_{ijk} + \sum_{j=1}^{m}\sum_{k=1}^{q}FP_{ijk}}$ & specificity from the user $i$\\
$acc_i = \sum_{j=1}^{m}\sum_{k=1}^{q}TP_{ijk} + \sum_{j=1}^{m}\sum_{k=1}^{q}TN_{ijk} / $ & accuracy from the user $i$\\
$(\sum_{j=1}^{m}\sum_{k=1}^{q}TP_{ijk} + \sum_{j=1}^{m}\sum_{k=1}^{q}FN_{ijk}+ $ & \\
$ \sum_{j=1}^{m}\sum_{k=1}^{q}TN_{ijk} + \sum_{j=1}^{m}\sum_{k=1}^{q}FP_{ijk} )$ & \\
$TP_{ijk} = \{0,1\}$ &  1 if the $k^{th}$ point is a true positive. \\
$TN_{ijk}= \{0,1\}$ &  1 if the $k^{th}$ point is a true negative. \\
$FP_{ijk}= \{0,1\}$ &  1 if the $k^{th}$ point is a false positive. \\
$FN_{ijk}= \{0,1\}$ &  1 if the $k^{th}$ point is a false negative. \\

$b_{0}$	&	intercept in the regression model\\
$b_{1}$	&	slope in the regression model \\
$u_j$	&	spatial component (CAR prior) on the image/location $j$ \\

$\mu_j$	&	mean of the beta distribution \\
$\phi$	&	precision parameter in the beta distribution \\
$x$	&	a predictor or disturbances\\
$\tau$ & precision parameter\\ 
$D$ & $m \times m$ diagonal matrix\\
$W$ & $m \times m$ adjacency matrix\\ 
\hline
\end{tabular}
}
\end{table}

\bibliographystyle{rss}
\bibliography{ref}

\clearpage
\section*{Supplementary materials}

\section{Example of a beta regression in presence of misclassification}
\label{sec:appp}

In this example, we illustrate how misclassification in the response variable produces biased regression coefficients. 
We simulate a dataset comprising 200 observations with a covariate ($x \sim \textrm{Uni}\left(0,1\right)$) and a response variable  that is beta distributed ($y \sim \textrm{Beta}\left(\alpha, \beta \right)$ ). 
The response $y$ is not directly observed but obtained from the elicited variable $\hat{y}$ produced from image classification. 
Consider four users ($i = 1, \cdots,4$ ) with different performance measures being $se_{i} = sp_{i} = \{0.90, 0.80, 0.70, 0.60\}$. 

\begin{lstlisting}[language=R]
# R codes
set.seed(201901)
n <- 200             # sample size
x <- runif(n, 0, 1)  # a covariate
beta <- c(-5, 10)     # fixing regression coefficients

mu <- 1 / (1 + exp(-(beta[1] + beta[2] * x))) # mean
phi = 50             # precision
a = mu * phi         # shape 1
b = -mu * phi + phi  # shape 2 
y = rbeta(n, a, b)   # reponse variable

df <- data.frame(x = x, y = y)

annot <- rep( c('cit1','cit2','cit3','cit4'), each = 50) 
df$annot <- sample(annot, size = nrow(df), replace = FALSE) 

perf <- data.frame(annot = c('cit1', 'cit2', 'cit3', 'cit4'),
 se = c(0.90, 0.80, 0.70, 0.60)) 
df <- merge(df, perf, by = 'annot')
df$yhat <- (df$y * df$se + (1 - df$y) * (1 - df$se)) 
# the apparent prop (apparent coral cover). 
# For simplicity let us assume se = sp

\end{lstlisting}

If the latent variable $y$ were observed directly we can retrieve the fixed regression coefficients well (Table \ref{table:tab1}) 
 
\begin{lstlisting}[language=R]
library("betareg")
m <- betareg(y ~ x, data = df) # model for the true fraction
summary(m)$coefficients$mean
\end{lstlisting}

\begin{table}
\caption{\label{table:tab1} Regression model, if the latent variable were observed directly or not there is no misclassification}
\begin{tabular}{rrrrr} 
  \hline
 & Estimate & Std. Error & z value & Pr($>$$|$z$|$) \\ 
  \hline
	(Intercept) & \textcolor{green}{-4.96275} & 0.09144 & -54.27365 & 0.00000 \\ 
  x & \textcolor{green}{9.92043} & 0.16941 & 58.55978 & 0.00000 \\ 
   \hline
	
\end{tabular}
\end{table}

\clearpage

\subsection*{Ignoring the misclassification problem:}
We obtain very poor estimates when the regression is performed directly from the observed apparent variable $\hat{y}$ (Table \ref{tab:ign}). 

\begin{lstlisting}[language=R]
m1 <- betareg(yhat ~ x, data = DF) # model for the apparent fraction
(sum1 <- summary(m1)$coefficients$mean)
\end{lstlisting}

\begin{table}
\caption{\label{tab:ign}Ignoring the misclassification}
\begin{tabular}{rrrrr} 
  \hline
	 & Estimate & Std. Error & z value & Pr($>$$|$z$|$) \\ 
  \hline
(Intercept) & \textcolor{red}{-1.47758} & 0.06734 & -21.94303 & 0.00000 \\ 
  x & \textcolor{red}{2.91030} & 0.11394 & 25.54236 & 0.00000 \\ 
   \hline
\end{tabular}
\end{table}

\subsection*{Accounting for misclassification using weights}
Using weights proportional to the performance measures yields slightly better estimates compared to the previous model but still very biased intercept and slope estimates (Table \ref{tab:usingW}).
 
\begin{lstlisting}[language=R]
m2 <- betareg(yhat ~ x, data = df, weights = acc) # model for the apparent fraction
sum2 <- summary(m2)$coefficients$mean
\end{lstlisting}

\begin{table}
\caption{\label{tab:usingW} Accounting for misclassification using weights}
\begin{tabular}{rrrrr} 
  \hline
	 & Estimate & Std. Error & z value & Pr($>$$|$z$|$) \\ 
  \hline
	(Intercept) & \textcolor{red}{-1.58172} & 0.07937 & -19.92916 & 0.00000 \\ 
  x & \textcolor{red}{3.11236} & 0.13495 & 23.06243 & 0.00000 \\ 
   \hline
\end{tabular}
\end{table}

\begin{figure}[h]
	\centering
		\includegraphics[width=4in]{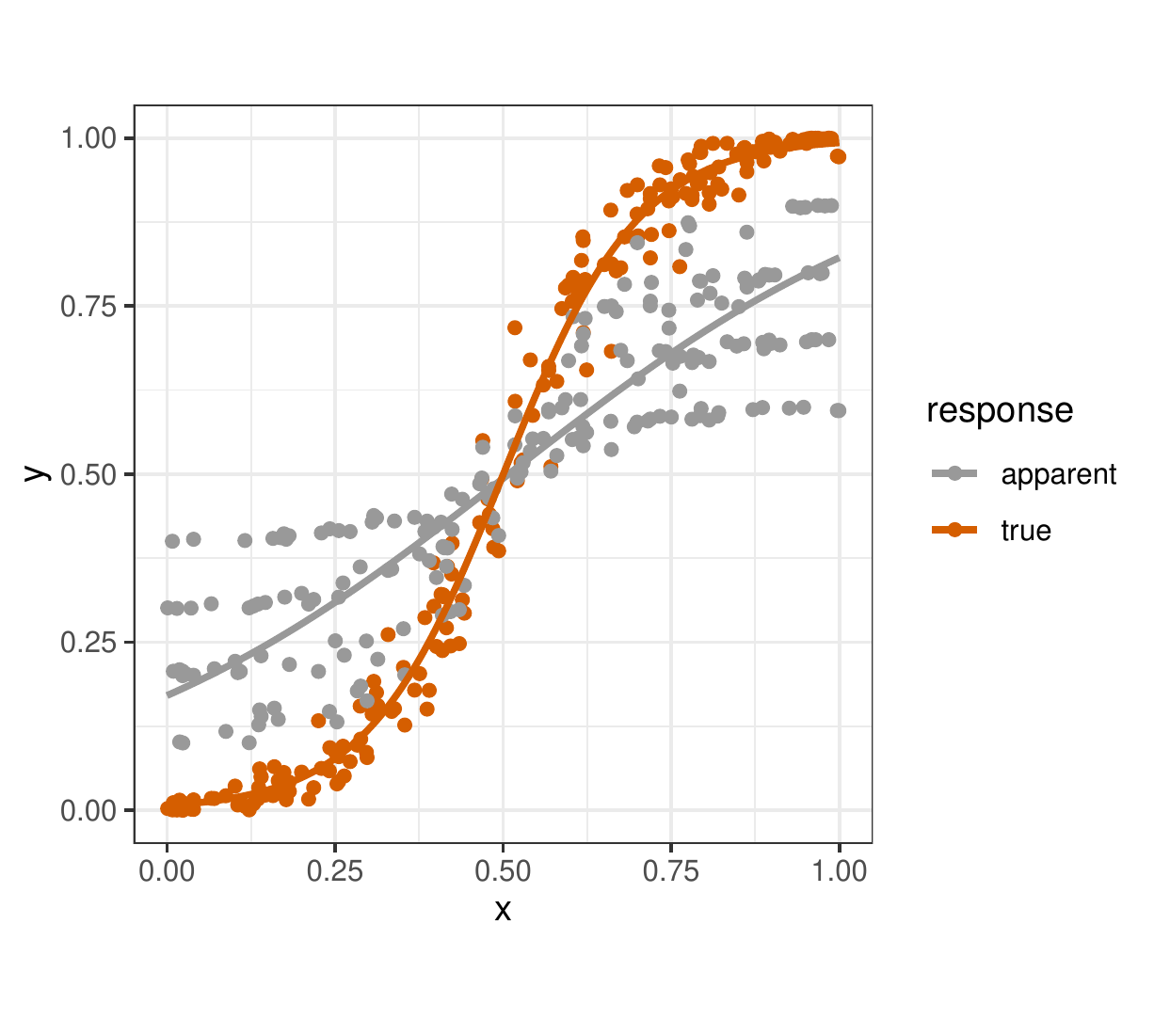}
	\caption{True unobserved response variable (in gray) and the observed apparent response (in orange) as a function of a predictor $x$.}
	\label{fig:FigAppendix}
\end{figure}

 Fig\ref{fig:FigAppendix} shows the unobserved true response variable (in gray) and the observed apparent response (in orange).

\pagebreak

\section{Posterior densities and trace plots from the simulation study}
\label{sec:posterior_plots}

\begin{figure}[htbp] 
	\centering
		\includegraphics[width=4.0in]{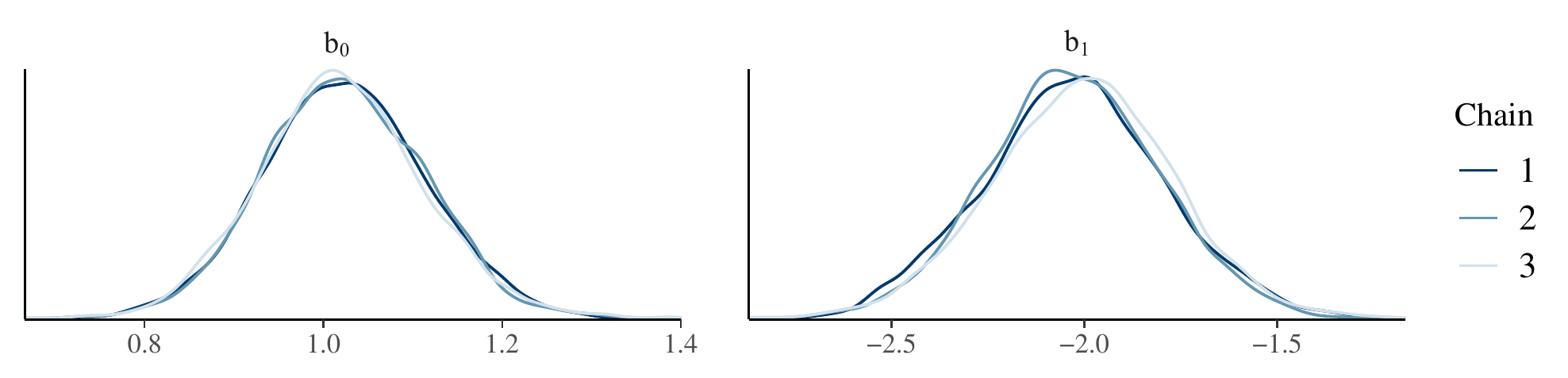}
	\caption{Posterior densities of the regression coefficients.} 
\label{fig:7Fig7}
\end{figure}

\begin{figure}[htbp]
	\centering
		\includegraphics[width=4.0in]{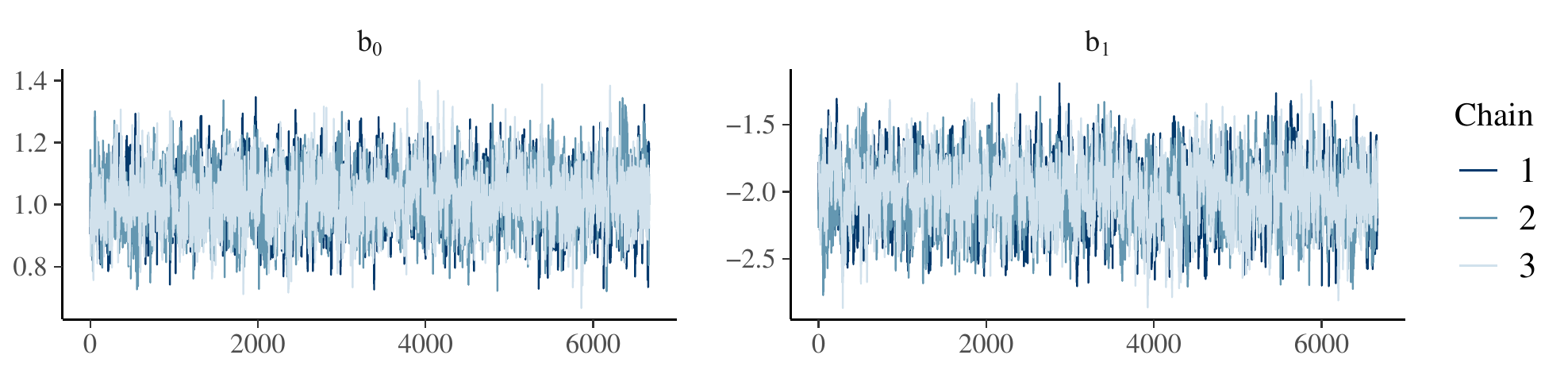}
	\caption{Trace plots of the regression coefficients.} 
\label{fig:8Fig8}
\end{figure}

\begin{figure}[htbp]
	\centering
		\includegraphics[width=4.0in]{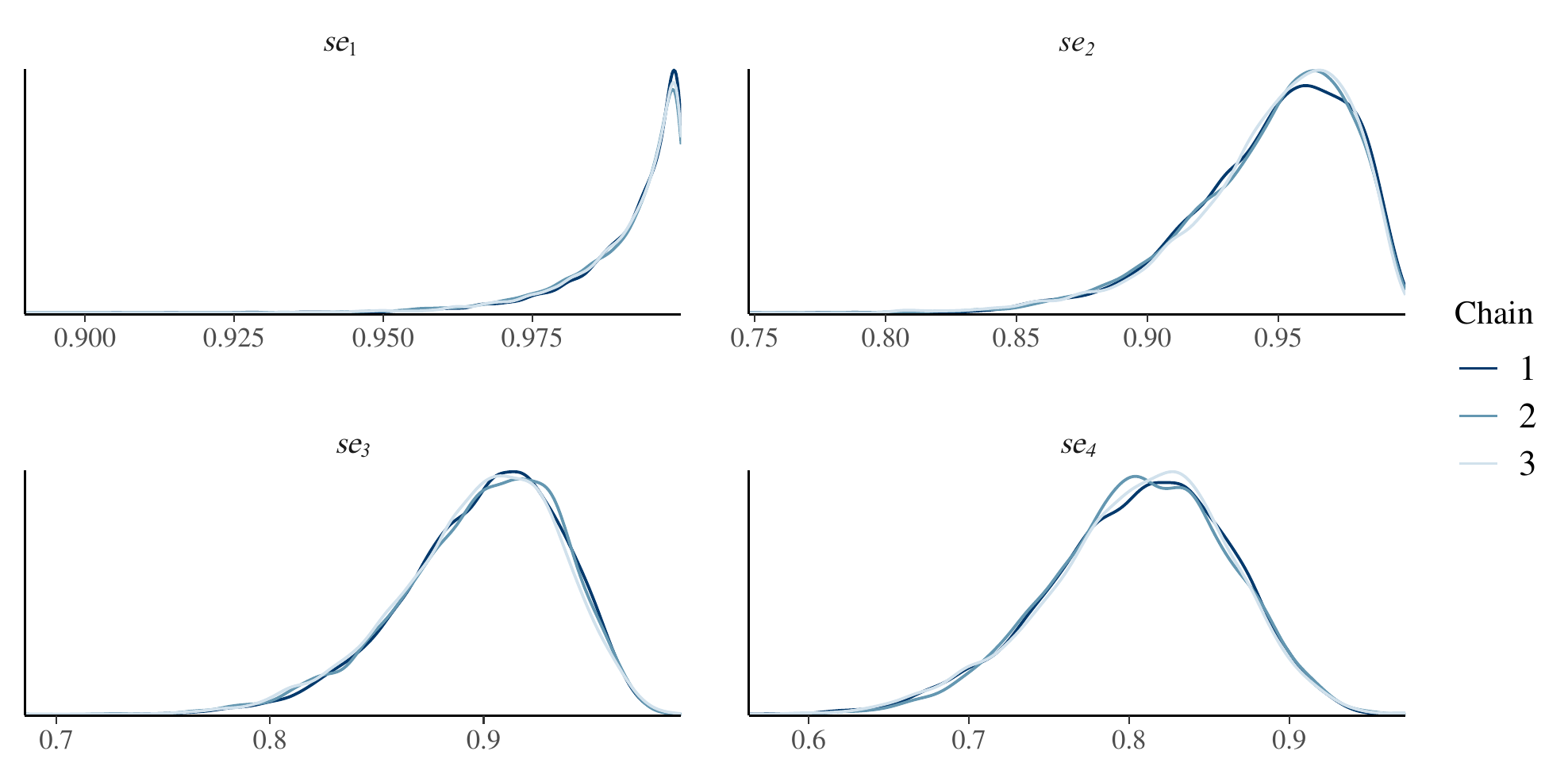}
                \caption{Posterior densities of the first four subjects' $se$. The true fixed values are: $se_1 = 0.99$,
                  $se_2 = 0.95$, $se_3 = 0.90$, $se_4 = 0.80$.}
\label{fig:9Fig9}
\end{figure}

\begin{figure}[htbp]
	\centering
		\includegraphics[width=4.0in]{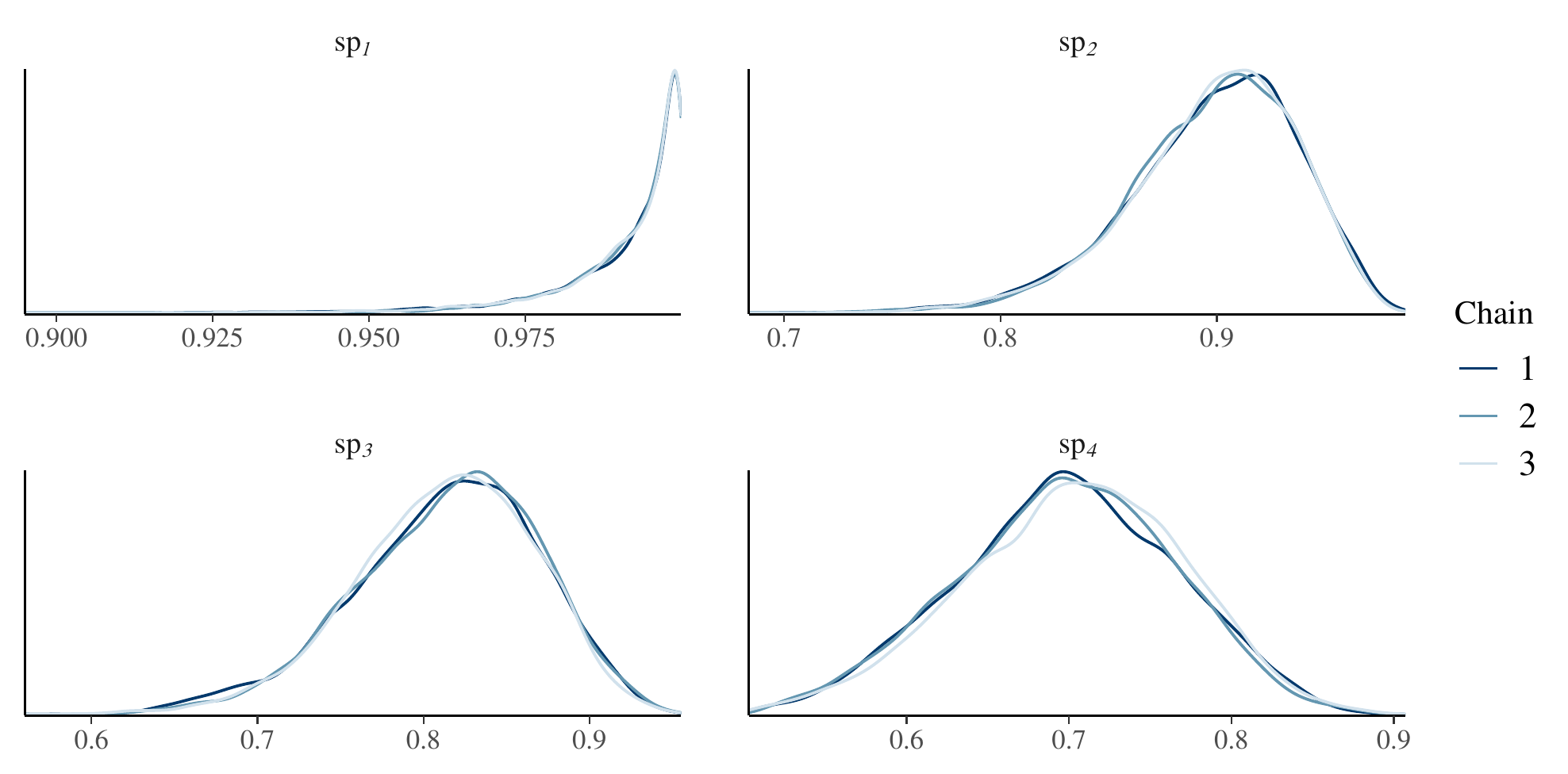}
	\caption{Posterior densities of the first four subjects' $sp$.   
	The true values are: $sp_1 = 0.99$, $sp_2 = 0.90$, $sp_3 = 0.80$, $sp_4 = 0.70$.} 
\label{fig:10Fig10}
\end{figure}

\end{document}